\documentclass[aps,preprint,prd,showpacs,showkeys,nofootinbib,tightenlines]{revtex4-1}

\usepackage{graphicx}  
\usepackage{amsmath}
\usepackage{bm}  
\usepackage{ulem}
\usepackage{amssymb}
\usepackage{comment}
\usepackage{lineno}

\newcommand{\bea}{\begin{eqnarray}}
\newcommand{\eea}{\end{eqnarray}}
\newcommand{\beq}{\begin{equation}}
\newcommand{\eeq}{\end{equation}}
\newcommand{\bqa}{\begin{eqnarray}}
\newcommand{\eqa}{\end{eqnarray}}

\begin{document}

\title{
Charm-meson Triangle Singularity\\
 in $\bm{e^+e^-}$ Annihilation into $\bm{ D^{*0}  \bar{D}^0 + \gamma }$}

\author{Eric Braaten}
\email{braaten.1@osu.edu}
\affiliation{Department of Physics,
         The Ohio State University, Columbus, OH\ 43210, USA}

\author{Li-Ping He}
\email{he.1011@buckeyemail.osu.edu}
\affiliation{Department of Physics,
         The Ohio State University, Columbus, OH\ 43210, USA}

\author{Kevin Ingles}
\email{ingles.27@buckeyemail.osu.edu}
\affiliation{Department of Physics,
         The Ohio State University, Columbus, OH\ 43210, USA}

\author{Jun Jiang}
\email{jiangjun87@sdu.edu.cn}
\affiliation{School of Physics, Shandong University, Jinan, Shandong 250100, China}

\date{\today}

\begin{abstract}
We calculate the cross section for $e^+ e^-$ annihilation into  $D^{*0} \bar D^0 +\gamma$
at center-of-mass energies near the $D^{*0} \bar D^{*0}$ threshold
under the assumption that $X(3872)$ is a weakly bound charm meson molecule.
The Dalitz plot has a $\bar D^{*0}$  resonance band in the squared  invariant mass  $t$ of $\bar D^0 \gamma$.
In the limit as the decay width of the $D^{*0}$ goes to 0,
the Dalitz plot also has a narrow band in the squared  invariant mass $u$ of $D^{*0} \bar D^0$
from a charm-meson triangle singularity.
At the physical value of the $D^{*0}$ width, the narrow band reduces to a shoulder.
Thus the triangle singularity cannot be observed directly as a peak in a differential cross section as a function of $u$.
It may however be observed indirectly as a local minimum in the $t$ distribution for events with $u$ 
below the triangle singularity.  The minimum is produced by the Schmid cancellation 
between triangle loop diagrams and  a tree diagram.
The observation of this  minimum would support the identification of $X(3872)$
 as a weakly bound charm meson molecule.
\end{abstract}

\smallskip
\maketitle

\section{Introduction}
\label{sec:Introduction}

Since early in this century, a large number of exotic hadrons whose constituents include a heavy quark and its antiquark
have been discovered in high energy physics experiments 
\cite{Ali:2017jda,Olsen:2017bmm,Karliner:2017qhf,Yuan:2018inv,Brambilla:2019esw}. 
The first of these exotic heavy hadrons to be discovered was the $X(3872)$ meson in 2003 \cite{Choi:2003ue}. 
Its $J^{PC}$ quantum numbers are $1^{++}$ \cite{Aaij:2013zoa}.
Its mass  is extremely close to the $D^{*0} \bar D^0$  threshold,
with the difference being less than about  0.2~MeV \cite{Tanabashi:2018oca}.
These results suggest that $X$ is a weakly bound S-wave charm-meson molecule
with the flavor structure
\begin{equation}
\big| X(3872) \rangle = \frac{1}{\sqrt2}
\Big( \big| D^{*0} \bar D^0 \big\rangle +  \big| D^0 \bar D^{*0}  \big\rangle \Big).
\label{Xflavor}
\end{equation}
There  are alternative models for $X$,
including a compact tetraquark state with constituents $c \bar c  q \bar q$
and the $\chi_{c1}(2P)$ charmonium state with constituents $c \bar c$
\cite{Ali:2017jda,Olsen:2017bmm,Karliner:2017qhf,Yuan:2018inv,Brambilla:2019esw}.
The nature of $X$ may have important implications for other exotic heavy hadrons.
One might have hoped the nature of  $X$ could be revealed by its decays.
The $X$ has been observed in 7 different decay modes, more than any other exotic heavy hadron.
Despite these many decay modes, a consensus on the nature of $X$ has not been achieved.

There may be aspects of the production of $X$ that are more effective at discriminating
between models than the decays of $X$.
If $X$ is a weakly bound charm-meson molecule, it can be produced by any reaction that can 
produce its constituents $D^{*0} \bar D^0$ and $D^0 \bar D^{*0}$. 
These constituents can be created at short distances of order $1/m_\pi$, where $m_\pi$ is the pion mass,
and they can subsequently be bound into $X$ at longer distances.
The $X$ can also be produced by the  creation of 
$D^{*0} \bar D^{*0}$, $D^{*0} D^{*-}$, or $D^{*+} \bar D^{*0}$ at short distances 
followed by the rescattering of the charm-meson pair into $X$ and a pion 
at longer distances \cite{Braaten:2019yua,Braaten:2019sxh}.

One way in which the nature of a hadron can be revealed by its production is through {\it triangle singularities}.
Triangle singularities are kinematic singularities that arise if three virtual particles that form a triangle 
in a Feynman diagram can all be on their mass shells simultaneously \cite{Karplus:1958zz,Landau:1959}.
There have been several previous investigations of the effects of triangle singularities on the production of
exotic heavy mesons \cite{Szczepaniak:2015eza,Liu:2015taa,Szczepaniak:2015hya,Guo:2017wzr}.
Guo recently pointed out that  any high-energy process 
that can create $D^{*0} \bar{D}^{*0}$ at short distances in an S-wave channel
will produce $X+ \gamma$ with a narrow peak near the $D^{*0} \bar{D}^{*0}$ threshold 
due to a charm-meson triangle singularity \cite{Guo:2019qcn}. 
Any high-energy process that can create an S-wave $D^* \bar{D}^*$ pair at short distances can also produce $X \pi$ 
with a narrow peak near the $D^* \bar{D}^*$ threshold due to a charm-meson triangle singularity.
Such a narrow peak arises in the production of $X \pi$ at hadron colliders \cite{Braaten:2019sxh}
and in decays of $B$ mesons  into $K+ X \pi$ \cite{Braaten:2019yua}.
It has been suggested that in the decay $B  \to X+K\pi$, $X$ is simply a peak 
from a triangle singularity \cite{Nakamura:2019nwd}.
The effects of the triangle singularity on the reaction $B^- \to K^-+ X\pi^0$ 
has recently been reexamined \cite{Sakai:2020ucu}.
The contribution of a triangle singularity to the decay
$X \to \pi^+\pi^-\pi^0$ has recently been studied \cite{Molina:2020kyu}.
Guo, Liu, and Sakai have  presented a review of triangle singularities in hadronic reactions 
with an emphasis on exotic heavy hadrons \cite{Guo:2019twa}.

The quantum numbers $1^{++}$ of $X$ allow $X+ \gamma$ to be produced by 
$e^+ e^-$ annihilation into a virtual photon.  The virtual photon can create
$D^{*0} \bar{D}^{*0}$ at short distances in a P-wave channel, and the charm-meson pair
can subequently rescatter into $X+ \gamma$.
The production of $X +\gamma$ in $e^+ e^-$ annihilation 
was first discussed by Dubynskiy and Voloshin  \cite{Dubynskiy:2006cj}.
They calculated the absorptive contribution to the cross section from 
$e^+ e^-$ annihilation into on-shell charm mesons $D^{*0} \bar{D}^{*0}$ 
followed by their rescattering into $X +\gamma$.
The absorptive part of the matrix element was calculated in the limit of zero decay width of the $D^{*0}$. 
Dubynskiy and Voloshin predicted that the cross section has a narrow peak  a few MeV 
above the $D^{*0} \bar D^{*0}$ threshold whose position and width depend on the $X$ binding energy.
As pointed out in Ref.~\cite{Braaten:2019gfj}, the narrow peak comes from a triangle singularity.
In the limit as the $X$ binding energy and the $D^{*0}$ width both go to zero,
there is a term in the matrix element that diverges logarithmically at a center-of-mass energy near the predicted peak.
Thus the $D^{*0}$ width may be as important as the $X$ binding energy
in determining the shape of the narrow peak.
The cross section for  $e^+ e^- \to X +\gamma$ in the energy region
near the $D^{*0} \bar{D}^{*0}$ threshold was calculated in Refs.~\cite{Braaten:2019gfj,Braaten:2019gwc},
taking into account the $D^{*0}$ width and including the dispersive as well as the absorptive contributions. 
For the physical $D^{*0}$ width, the position of the narrow peak is fairly insensitive to the $X$ binding energy.
The width of the peak is determined by both the binding energy and the $D^{*0}$ width.

Since $D^{*0} \bar D^0$ are constituents of $X$, the annihilation of  $e^+ e^-$ can also produce $D^{*0} \bar D^0+ \gamma$.
The matrix element for this reaction is the sum of a tree diagram and a pair of loop diagrams with a charm-meson triangle.
In the limit of zero $D^{*0}$ width, the differential cross section
has a double-logarithmic divergence as a function of the $D^{*0} \bar D^0$ invariant mass
 in a narrow range of the center-of-mass energy.
An interesting aspect of this reaction is a cancellation pointed out by Schmid \cite{Schmid:1967ojm}.
In the differential cross section integrated over  the $\bar D^0 \gamma$ invariant mass,
the double-logarithmic divergence is canceled by
the interference between the tree diagram and the loop diagrams.
Anisovitch and Anisovitch showed that the cancellation leaves a single-logarithmic divergence \cite{Anisovich:1995ab}.

In this paper, we  calculate the cross section for $e^+ e^- \to D^{*0} \bar D^0+ \gamma$
at center-of-mass energies near the $D^{*0} \bar{D}^{*0}$ threshold.
The Schmid cancellation of the leading divergence from the triangle singularity
is most conveniently revealed by expressing the cross section in a Lorentz-invariant form.
In Section~\ref{sec:D*D*}, we repeat the calculation in Ref.~\cite{Braaten:2019gwc}
of the cross section for  $e^+e^- \to D^{*0} \bar D^{*0}$ near the threshold using Lorentz-invariant variables. 
In Section~\ref{sec:Xgamma}, we repeat the calculation in Ref.~\cite{Braaten:2019gwc}
 of the cross section for  $e^+e^-\to X+ \gamma$ near the $D^{*0} \bar D^{*0}$ threshold using Lorentz-invariant variables. 
In Section~\ref{sec:DDgamma}, we calculate the cross section for  $e^+e^-\to D^{*0} \bar D^0+ \gamma$
 and identify the triangle singularity.
In Section~\ref{sec:Schmid}, we demonstrate the Schmid cancellation  of the leading logarithm
from that singularity.
We point out that the triangle singularity can be observed indirectly through a local minimum 
in the differential cross section produced by interference associated with the Schmid cancellation.
Our results are summarized in Section~\ref{sec:Summary}.
The nonrelativistic result for the loop amplitude with the triangle singularity is reproduced in Appendix~\ref{LoopIntegralNR}.
The loop integral that produces the triangle singularity is calculated
in terms of Lorentz-invariant variables in Appendix~\ref{LoopIntegralR}.


\section{Production of   $\bm{D^{*0} \bar{D}^{*0}}$ near  threshold}
\label{sec:D*D*}

\begin{figure}[hbt]
\includegraphics*[width=0.35\linewidth]{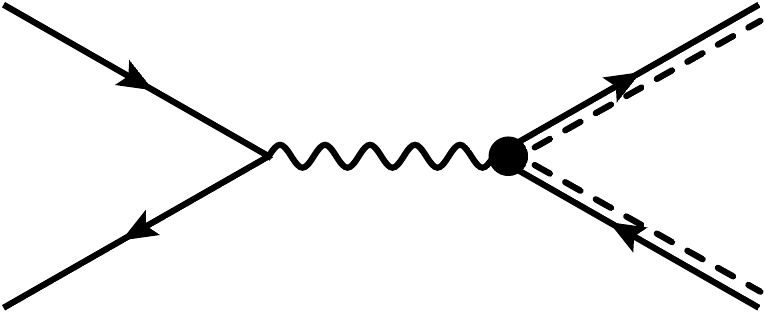} 
\caption{Feynman diagram for $e^+e^-\to  D^{*0} \bar D^{*0}$. 
The spin-1 charm mesons $D^{*0}$ and $\bar D^{*0}$ are represented by double lines 
consisting of a dashed line and a solid line with an arrow.
}
\label{fig:eetoD*D*}
\end{figure}

A pair of spin-1 charm mesons $D^{*0} \bar D^{*0}$ can be produced from the annihilation of $e^+e^-$ 
into a virtual photon. The Feynman diagram for this process is shown in Fig.~\ref{fig:eetoD*D*}.
The cross section for this reaction near the threshold was calculated in Ref.~\cite{Braaten:2019gwc} using 
nonrelativistic approximations for the charm mesons. 
We repeat the calculation here using  a relativistic formalism.

The matrix element for $e^+e^-$ with total momentum $Q$ to produce 
$D^{*0}$ and $\bar D^{*0}$ with 4-momenta $p$ and  $\bar p$ has the form
\begin{equation}
\mathcal{M}=  i\frac{e^2(2M_{*0})}{s}\, \bar v \gamma_\mu u\,
\mathcal{A}(Q)^{\mu\nu\lambda\sigma} p_\nu\, \varepsilon^*(p)_\lambda\, \varepsilon^*(\bar p)_\sigma,
\label{MeetoD*D*}
\end{equation}
where $\sqrt{s}$ is the center-of-mass energy and $\bar v$ and $u$ are the spinors for the colliding $e^+$ and $e^-$.
The factor of $2M_{*0}$, where $M_{*0}$ is the mass of the  $D^{*0}$, 
compensates for the relativistic normalization of charm meson states.
Near the threshold $s= 4 M_{*0}^2$ for producing $D^{*0} \bar D^{*0}$, 
the charm-meson pair is produced in a P-wave state with total spin either 0 or 2.
At energies close enough to threshold that higher partial waves can be ignored,
 the Lorentz tensor $\mathcal{A}(Q)^{\mu\nu\lambda\sigma}$  in Eq.~\eqref{MeetoD*D*} is
\begin{equation}
\mathcal{A}(Q)^{\mu\nu\lambda\sigma}= A_0 \, g_Q^{\mu\nu} g_Q^{\lambda\sigma} 
+ \frac{3}{2\sqrt{5}} A_2 \left(g_Q^{\mu \lambda} g_Q^{\nu \sigma} + g_Q^{\mu \sigma} g_Q^{\nu \lambda} 
- \frac{2s^2}{s^2+32 M_{*0}^4} g_Q^{\mu \nu} g_Q^{\lambda \sigma}  \right) ,
\label{A[eetoD*D*]}
\end{equation}
where $g_Q^{\mu \nu} = g^{\mu \nu}-Q^\mu Q^\nu/Q^2$. 
The coefficients  $A_0$ and $A_2$ are  the amplitudes for creating $D^{*0} \bar D^{*0}$ 
in a P-wave channel with total spin 0 and 2, respectively.  
The $s$-dependence of the last term in Eq.~\eqref{A[eetoD*D*]} is necessary
to eliminate interference between the two amplitudes in the differential cross section.
The matrix element in Eq.~\eqref{MeetoD*D*} with energy-dependent amplitudes $A_0(s)$ and $A_2(s)$
is a good approximation at energies less than about 100~MeV above the threshold \cite{Braaten:2019gwc}.
That matrix element with constant amplitudes $A_0$ and $A_2$
is a good approximation at energies less than about 10~MeV above the threshold \cite{Braaten:2019gwc}.
In this region, $s-4M_{*0}^2$ is numerically smaller than $4\delta^2$,
where $\delta = 142.0$~MeV is the mass difference between $D^{*0}$ and $D^0$.
We can therefore consider $s-4M_{*0}^2$ to be order $\delta^2$.

The  differential cross section for producing  $D^{*0} \bar D^{*0}$ with scattering angle $\theta$ 
in the center-of-momentum frame is 
\begin{eqnarray}
\frac{d\sigma}{d\Omega} &=& \frac{\alpha^2 (s^2 + 32 M_{*0}^4)}{128 M_{*0} ^2 s} 
\left(\frac{s-4M_{*0}^2}{s} \right)^{3/2}
\left[ \left( |A_0|^2 +  \frac{288 M_{*0}^4 s^2}{5(s^2 + 32 M_{*0}^4)^2}  |A_2|^2 \right) (1 - \cos^2\theta) \right.
\nonumber\\
&& \hspace{6.3cm} \left.
+ \frac{18 M_{*0}^2 s}{5(s^2 + 32 M_{*0}^4)} |A_2|^2 \, (1 + \cos^2\theta) \right] .
\label{dsigma/dOmega}
\end{eqnarray}
In the region where $s-4M_{*0}^2$ is order $\delta^2$,
we can  simplify the cross section with errors of order $(\delta/M_{*0})^2$
by setting $s=4M_{*0}^2$  everywhere except in the factors of $(s-4M_{*0}^2)/s$.
The  cross section  reduces to
\begin{equation}
\frac{d\sigma}{d\Omega} = \frac{3\alpha^2}{32} \left(\frac{s-4M_{*0}^2}{s} \right)^{3/2}
\left[ \left( |A_0|^2 +  \frac{2}{5}  |A_2|^2 \right) (1 - \cos^2\theta) 
+ \frac{3}{10} |A_2|^2 \, (1 + \cos^2\theta) \right]   .
\label{dsigma/dOmega-th}
\end{equation}
This agrees with the nonrelativistic cross section calculated in Ref.~\cite{Braaten:2019gwc}.
The cross section integrated over angles is
\begin{equation}
\sigma[ e^+ e^- \to D^{*0} \bar D^{*0}] = \frac{\pi \alpha^2}{4} \left(\frac{s-4M_{*0}^2}{s} \right)^{3/2}
\left[ |A_0|^2  + |A_2|^2  \right]   .
\label{sigma-threshold}
\end{equation}

\begin{figure}[hbt]
\includegraphics*[width=0.7\linewidth]{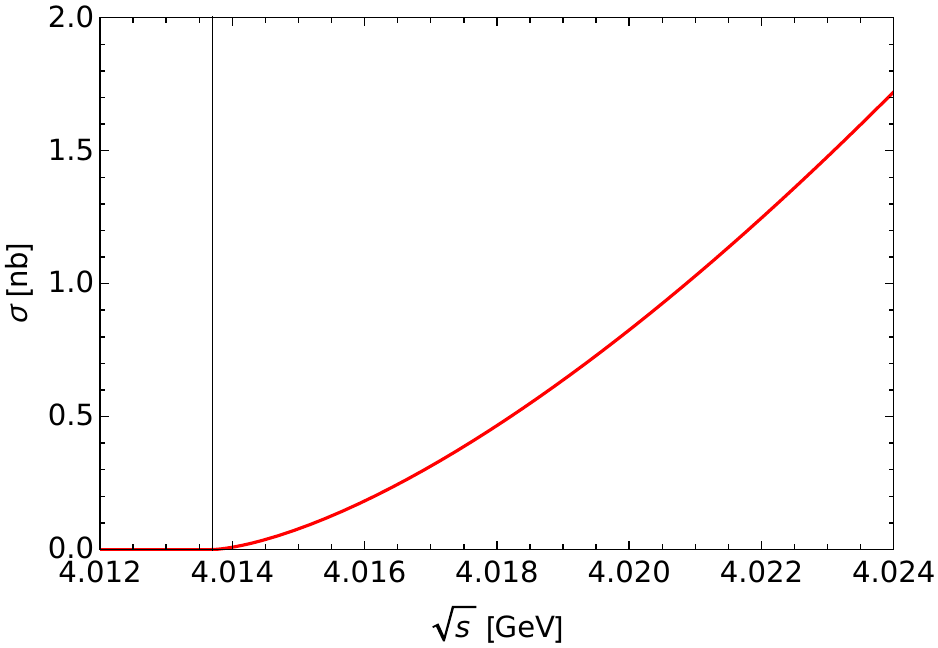} 
\caption{
Cross section for producing $D^{*0} \bar D^{*0}$ as a function of 
the center-of-mass energy $\sqrt{s}$.  The values of $|A_0|$ and $|A_2|$ are given  in Eq.~\eqref{A0,A2}. 
The vertical line is the $D^{*0} \bar D^{*0}$ threshold.
 }
\label{fig:sigmaD*D*}
\end{figure}

The constant factor $|A_0|^2  + |A_2|^2$ in Eq.~\eqref{sigma-threshold} can be determined experimentally by measuring the 
cross section at center-of-mass energies $\sqrt{s}$ within about 10~MeV of the $D^{*0} \bar D^{*0}$ threshold at 4013.7~MeV.
The values of $|A_0|$ and $|A_2|$ can be determined separately by measuring the angular distribution 
and fitting it to the expression in Eq.~\eqref{dsigma/dOmega-th}.
Estimates of the amplitudes $|A_0|$  and $|A_2|$ were presented in Ref.~\cite{Braaten:2019gfj}:
\begin{equation}
|A_0| = 8~\mathrm{GeV}^{-1}, \qquad |A_2| = 15~\mathrm{GeV}^{-1} .
\label{A0,A2}
\end{equation}
The cross section  for $D^{*0} \bar D^{*0}$ predicted using  Eq.~\eqref{sigma-threshold} with these amplitudes
is shown in Fig.~\ref{fig:sigmaD*D*} for center-of-mass energies up to about 10~MeV above the threshold.
The cross section is large enough that it should be relatively easy for the BESIII collaboration 
to measure $|A_0|$ and $|A_2|$.

The amplitudes $|A_0|$  and $|A_2|$ in Eq.~\eqref{A0,A2} were actually obtained by fitting
cross sections for $e^+ e^-$ annihilation into $D^{*+} D^{*-}$ to Eq.~\eqref{sigma-threshold}
with $M_{*0}$ replaced by the mass $M_{*+}$ of $D^{*+}$.
The justification for approximating the amplitudes for producing  $D^{*0} \bar D^{*0}$ by those for producing 
$D^{*+} D^{*-}$  is that both cross sections are dominated near the threshold by the isospin-0 charmonium resonance 
$\psi(4040)$. The amplitudes $|A_0|$  and $|A_2|$ were 
determined in Ref.~\cite{Braaten:2019gfj} by fitting cross sections obtained by
Uglov {\it et al.}\  \cite{Uglov:2016orr} by analyzing Belle data  on  $e^+ e^-$ annihilation 
into pairs of charm mesons  \cite{Abe:2006fj,Pakhlova:2008zza}. 
The Belle data was also analyzed by Du {\it et al.}\  \cite{Du:2016qcr}.
Their fit gives a cross section for $D^{*+} D^{*-}$
at $\sqrt{s} = 4.040$~GeV that is only 6\% smaller than the fit of Uglov {\it et al.}.
However their ratio 0.81 of the spin-2 and spin-0 cross sections at that energy
is  significantly smaller   than the ratio  2.92 from the fit  of Uglov {\it et al.}. 
This suggests that the value $|A_0|^2  + |A_2|^2= 290~\mathrm{GeV}^{-2}$
from Eq.~\eqref{A0,A2} may be considerably more accurate than the value of the ratio  $|A_2|/|A_0|$.


\section{Production of   $\bm{X+ \gamma}$ near  the $\bm{D^{*0} \bar{D}^{*0}}$ threshold}
\label{sec:Xgamma}

If the $X(3872)$ is a weakly bound charm-meson molecule, its constituents are
the superposition of charm mesons in Eq.~\eqref{Xflavor}.
The $D^{*0}$ width gives an important contribution to the decay width of $X$.
The full decay  width of $D^{*0}$ has not been measured.  It can be predicted from measurements of 
 the full decay width $\Gamma_{*+}$ of $D^{*+}$ and the decay branching fractions of $D^{*0}$  and $D^{*+}$
using chiral symmetry and isospin symmetry: 
\begin{equation}
\frac{\Gamma_{*0}}{\Gamma_{*+}}=  \frac12 \left(\frac{M_{*+}}{M_{*0}} \right)^{\!\!5} 
\left(\frac{\lambda(M_{*0}^2, M_0^2,m_0^2)}{\lambda(M_{*+}^2, M_0^2,m_+^2)}\right)^{\!\!3/2}
\frac{\mathrm{Br}[D^{*+} \to D^0\pi^+]}{\mathrm{Br}[D^{*0} \to D^0\pi^0]},
\label{Gamma*ratio}
\end{equation}
where $M_0$  is the mass of $D^0$, 
$m_0$ and $m_+$ are the masses of $\pi^0$ and  $\pi^+$,
and $\lambda(x,y,z) = x^2 + y^2 + z^2 -2(xy+yz+zx)$  is the K\"allen function.
The resulting prediction is $\Gamma_{*0} = (55.4 \pm 1.5)$~keV,
which agrees with  the result in Ref.~\cite{Guo:2019qcn}.
It is consistent within errors with the result in Ref.~\cite{Rosner:2013sha}.
The corresponding prediction for the radiative decay width of the $D^{*0}$ 
is $\Gamma[D^{*0} \to D^0 \gamma] = (19.6 \pm 0.9)$~keV.
This is consistent within errors with the result in Ref.~\cite{Rosner:2013sha}.
In this paper, we consider the $D^{*0}$ width to be accurately predicted,
and we take its value to be $\Gamma_{*0} = 55$~keV.

The present value of the difference $E_X$ between the mass of $X$ 
and the energy of the $D^{*0} \bar D^0$ scattering threshold is \cite{Tanabashi:2018oca}
\begin{equation}
E_X \equiv M_X - (M_{*0}\!+\!M_0) =( +0.01 \pm 0.18)~\mathrm{MeV}.
\label{EX-exp}
\end{equation}
The central value corresponds to a charm-meson pair just above the scattering threshold.
The value lower by $1\sigma$ corresponds to a bound state with binding energy $|E_X| =0.17$~MeV.  
The upper bound on $|E_X|$ with 90\% confidence level is 0.22~MeV.
Since the binding energy is so small, we will often use $M_X$ as a compact expression for $M_{*0}\!+\!M_0$.

We assume in this paper that $X$ is a narrow bound state.
The production of the $X$ resonance feature in the case where $X$ is 
 instead a virtual state was considered in the Appendix of Ref.~\cite{Braaten:2019gwc}
and in Ref.~\cite{Sakai:2020ucu}.
If $X$ is a narrow bound state, its energy $E_X$ and its decay width $\Gamma_X$
can be expressed in terms of the complex binding momentum $\gamma_X$ as
\begin{subequations}
\begin{eqnarray}
 E_X &=& -\frac{\mathrm{Re}[\gamma_X]^2 - \mathrm{Im}[\gamma_X]^2}{2\mu},
\label{EX-gammaX} 
 \\
\Gamma_X &=& \Gamma_{*0}  + \frac{2\, \mathrm{Re}[\gamma_X]\, \mathrm{Im}[\gamma_X]}{\mu},
\label{GammaX-gammaX} 
\end{eqnarray}
\label{EX,GammaX}%
\end{subequations}
where $\mu=M_{*0}M_0/(M_{*0}\!+\!M_0)$ is the reduced mass of $D^{*0} \bar D^0$.
The energy and decay width of the $X$ bound state should be distinguished from the energy and width
of the $X$ resonance feature, which has contributions from above the  $D^{*0} \bar D^0$ threshold
and may depend on the production process \cite{Braaten:2019ags}.
The first term in $\Gamma_X$ in Eq.~\eqref{GammaX-gammaX} 
can be interpreted as the partial decay width of the bound state into $D^0 \bar D^0 \pi^0$ and $D^0 \bar D^0 \gamma$.
The second term can be interpreted as its partial decay width into other decay modes, such as $J/\psi\,\pi^+\pi^-$.
To illustrate the possible dependence of production rates on the binding energy and decay width of $X$,
we will consider three choices for $E_X$, $\Gamma_X$ 
with the physical $D^{*0}$ width $\Gamma_{*0} = 55$~keV:
\begin{enumerate}
\item
$E_X = -0.05$~MeV, $\Gamma_X = \Gamma_{*0}$,
which corresponds to the real binding momentum $\gamma_X = 9.83$~MeV,
\item
$E_X = -0.10$~MeV, $\Gamma_X = \Gamma_{*0}$,
which corresponds to the real binding momentum $\gamma_X = 13.90$~MeV,
\item
$E_X = -0.05$~MeV, $\Gamma_X =2 \, \Gamma_{*0}$,
which corresponds to the complex binding momentum $\gamma_X = (10.17 +2.61i)$~MeV.
\end{enumerate}
The comparison of the first and second choices shows the effect of doubling the binding energy of $X$.
The comparison of the first and third choices shows the effect of doubling the width of $X$.
We will also sometimes consider zero decay width for $X$, 
which requires the unphysical $D^{*0}$ width $\Gamma_{*0} = 0$:
\begin{enumerate}
\setcounter{enumi}{3}
\item
$E_X = -0.05$~MeV, $\Gamma_X =0$, which corresponds to $\gamma_X = 9.83$~MeV,
\item
$E_X = 0$, $\Gamma_X =0$, which corresponds to $\gamma_X = 0$.
\end{enumerate}

\begin{figure}[hbt]
\includegraphics*[width=0.35\linewidth]{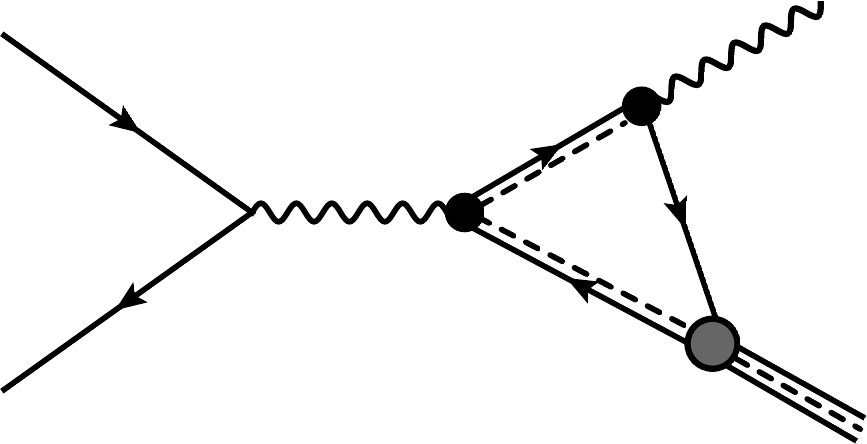} \quad
\includegraphics*[width=0.35\linewidth]{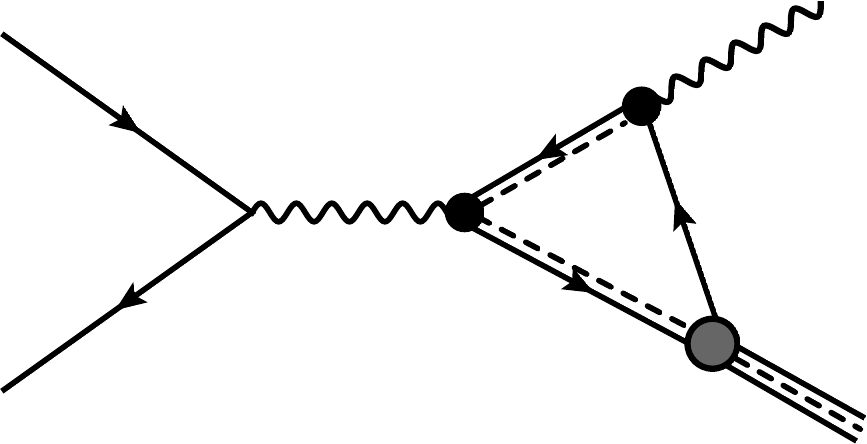} 
\caption{Feynman diagrams for  $e^+e^-\to X+\gamma$ from rescattering of  $D^{*0} \bar D^{*0}$. 
The $X$ is represented by a triple line consisting of two solid lines and a dashed line. 
The spin-0 charm mesons $D^0$ and $\bar D^0$ are represented by solid lines with an arrow.
}
\label{fig:eetogammaX}
\end{figure}

The $X$ can be produced in $e^+e^-$ annihilation through the creation of $D^{*0} \bar D^{*0}$
by a virtual photon followed by the rescattering of the charm-meson pair into $X+\gamma$.
The Feynman diagrams for this process are shown in Fig.~\ref{fig:eetogammaX}. 
The cross section for this reaction near the $D^{*0} \bar D^{*0}$ threshold was calculated in Ref.~\cite{Braaten:2019gwc} 
using nonrelativistic approximations for the charm mesons.
We repeat the calculation here using  a relativistic formalism.
We first describe each of the vertices in Fig.~\ref{fig:eetogammaX} that involve charm mesons.

At center-of-mass energies near the $D^{*0} \bar D^{*0}$ threshold,
the vertex for the virtual photon with momentum $Q=p+\bar p$ and Lorentz index $\mu$ to create $D^{*0}$ and $\bar D^{*0}$ 
with momenta $p$ and $\bar p$ and with  Lorentz indices $\lambda$ and $\sigma$ is 
$e(2M_{*0} ) \mathcal{A}(Q)^{\mu\nu\lambda\sigma} p_\nu$, where 
the  tensor $\mathcal{A}(Q)^{\mu\nu\lambda\sigma} $  is given in Eq.~\eqref{A[eetoD*D*]}.

The vertex for the radiative transition of $D^{*0}$
with momentum $p$ and index $\alpha$  to $D^0$ by emitting a photon with momentum $q$ and index $\beta$
is $\sqrt{4M_{*0} M_0}\, e\nu\,  \epsilon^{\alpha \beta\lambda\sigma}  (p_{\lambda}/M_{*0}) q_\sigma$.
The transition magnetic moment $e\nu$ can be determined from the radiative decay width of $D^{*0}$:
\begin{equation}
\Gamma[D^{*0} \to D^0 \gamma] =  \frac{\alpha \nu^2M_0(M_{*0}^2 -M_0^2)^3}{6 M_{*0}^4} .
\label{GammaD*0}
\end{equation}
Using the predicted radiative width $\Gamma[D^{*0} \to D^0 \gamma] = (19.6 \pm 0.9)$~keV,
the factor $\nu$ in the transition magnetic moment is determined to be  $\nu= 0.92~\mathrm{GeV}^{-1}$.

The binding of $D^{*0} \bar D^{0}$ or $D^{0} \bar D^{*0}$ into $X$ 
can be described within an effective field theory called XEFT \cite{Fleming:2007rp,Braaten:2015tga}.
The relativistic generalization of the vertex for the coupling of $D^{*0}\bar D^0$  or $D^{0}\bar D^{*0}$ to $X$ 
with momentum $P$ can be expressed as 
\begin{equation}
-i \sqrt{8M_X M_{*0} M_0}\, (\pi \gamma_X/\mu^2)^{1/2}\, (g^{\mu\nu} - P^\mu P^\nu/P^2),
\label{vertexX}
\end{equation}
where $\mu$ and $\nu$ are the Lorentz indices of the  spin-1 charm meson and $X$. 
In the rest frame of $X$, the vertex in Eq.~\eqref{vertexX} reduces to the nonrelativistic vertex
$i (\pi \gamma_X/\mu^2)^{1/2}\, \delta^{mn}$ \cite{Braaten:2010mg} multiplied by a factor $(8M_X M_{*0} M_0)^{1/2}$
that compensates for the relativistic normalization of states.

The matrix element for $e^+e^-\to X +\gamma$ is the sum of the two diagrams with  charm-meson loops
in Fig.~\ref{fig:eetogammaX}. 
The loop  integral produces a {\it triangle singularity} \cite{Braaten:2019gfj}.
The singularity arises from  the integration region where the three charm mesons whose  lines form triangles
in the diagrams in Fig.~\ref{fig:eetogammaX} are all on their mass shells simultaneously. 
The two charm mesons that become constituents of $X$ are both  on their mass shells 
in the limit where the binding energy is 0.  
The two spin-1 charm mesons can be both on their mass shells if the decay width $\Gamma_{*0}$ is 0.
The photon energy can be tuned so that the charm mesons before and after the radiative  transition 
are both on their mass shells.  
The  matrix element has a logarithmic branch point  from the triangle singularity at a complex value of $s$ 
that becomes real in the limit where $E_X$, $\Gamma_X$, and $\Gamma_{*0}$ are all 0.
That real branch  point is determined in Appendix~\ref{LoopIntegralR}:
\begin{equation}
s_\triangle = 4M_{*0}^2 + (M_{*0}/M_0) \, (M_{*0} - M_0)^2.
\label{s-triangle}
\end{equation}
It can be expressed more concisely as $s_\triangle = (M_{*0}/M_0) M_X^2$.
The predicted center-of-mass energy for the triangle singularity is $\sqrt{s_\triangle} = 4016.4$~MeV,
which is about 2.7~MeV above the $D^{*0} \bar D^{*0}$ threshold.
A nonrelativistic approximation for $s_\triangle$ was obtained in Ref.~\cite{Braaten:2019gwc}.
If we set $M_0 =M_{*0}- \delta$, $s_\triangle$ can be expanded in powers of $\delta$. 
The expansion of the nonrelativistic approximation for  
$s_\triangle$ agrees with the relativistic result 
in Eq.~\eqref{s-triangle} through third order in $\delta$, but it disagrees at fourth order. 

The nonrelativistic loop integral from the diagrams in Fig.~\ref{fig:eetogammaX}
was calculated in Ref.~\cite{Braaten:2019gwc}. 
The integral is ultraviolet convergent, and it can be reduced to a scalar function 
given analytically in  Appendix~\ref{LoopIntegralNR}.
The relativistic loop integral from the diagrams in Fig.~\ref{fig:eetogammaX} is quadratically ultraviolet divergent.  
The integral depends on the center-of-mass energy 
and the invariant mass $\sqrt{u}$ of the charm-meson-pair system that recoils against the photon.
In the region near the triangle singularity where $s-4M_{*0}^2$ and $u-M_X^2$ are order $\delta^2$,
the loop integral  can be expanded in powers of $\delta/M_{*0}$.
The leading terms in the expansion are ultraviolet finite.
In Appendix~\ref{LoopIntegralR},  the relativistic loop integral is reduced to the Lorentz-scalar function $F(s,u)$
defined by the ultraviolet-convergent momentum integral in Eq.~\eqref{FLorentz}.
For $s-4M_{*0}^2$ and $u-M_X^2$ both of order $\delta^2$,
the integral is dominated by  momentum scales of order $\delta$
and smaller, and the leading term scales as $1/(M_X \delta)$.
In the matrix element for producing $X+\gamma$, $F(s,u)$ is evaluated at 
$u=M_X^2 \!-\! i M_X\Gamma_X$, which is the position of the complex pole in the elastic scattering amplitude
for the charm mesons $D^{*0}$ and $\bar D^0$.

The matrix element  for $e^+e^-\to X +\gamma$ can be calculated  most conveniently by first reducing 
the charm-meson loops to an effective vertex for the coupling of the virtual photon 
to $X$ and a real photon given  in Eq.~\eqref{effvertex:Xgamma}.
The matrix element for producing $X$ with momentum $P$ and $\gamma$ with momentum $q$ is
\begin{equation}
\mathcal{M} = \frac{4e^3 \nu  (2M_X)^{3/2}  \sqrt{\pi \gamma_X}}{s} \, F(s,M_X^2 \!-\!i M_X\Gamma_X)\,
 \bar v \gamma_\mu u \, \mathcal{A}(Q)^{\mu \nu \lambda \sigma} q_\nu \, 
\epsilon_{\sigma\alpha \beta \tau}Q^\alpha q^\beta\, \varepsilon^*(P)_\lambda \, \varepsilon^*(q)^\tau.
\label{MeetoXgamma}
\end{equation}
The  differential cross section for producing $X + \gamma$ with  scattering angle $\theta$ is
\begin{eqnarray}
\frac{d\sigma}{d\Omega} &=&\frac{8 \pi^2 \alpha^3 \nu^2\,  |\gamma_X| \, M_X^3 (s-M_X^2)^5}{s^4}
 \left| F(s,M_X^2 \!-\!i M_X\Gamma_X)  \right|^2
\nonumber\\
&&
\times \left[ \left| A_0 - \frac{3 s^2}{\sqrt{5}\,(s^2+32M_{*0}^4)} A_2 \right|^2 ( 1- \cos^2 \theta ) 
+ \frac{9(s+M_X^2)^2}{160sM_X^2 }|A_2|^2 (1+ \cos^2 \theta)\right] .
\label{dsigma/dOmega:Xgamma}
\end{eqnarray}
In the region where $s-4M_{*0}^2$ is order $\delta^2$,
we can  simplify the cross section with errors of order $(\delta/M_{*0})^2$
by setting $s=4M_{*0}^2$ everywhere except in the factors of $(s-M_X^2)/s$
and in the function $F$.  The cross section reduces to
\begin{eqnarray}
\frac{d\sigma}{d\Omega} &=&
32\pi^2 \alpha^3 \nu^2\,  |\gamma_X| \, M_X^3 M_{*0}^2\left( \frac{s-M_X^2}{s} \right)^5
 \left|F(s,M_X^2 \!-\!i M_X\Gamma_X) \right|^2
\nonumber\\
&& \hspace{3cm}
\times \left[ \left| A_0 - \frac{1}{\sqrt{5}} A_2 \right|^2 ( 1- \cos^2 \theta ) 
+ \frac{9}{40}|A_2|^2 (1+ \cos^2 \theta)\right] ,
\label{dsigma/dOmega:Xgammathresh}
\end{eqnarray}
where $M_X=M_{*0} \!+\! M_0$ except in the second argument of the loop amplitude $F$,
which depends sensitively on $E_X$ through that argument.
The cross section for producing  $X+\gamma$ integrated over angles  is
\begin{eqnarray}
\sigma[e^+ e^- \to X \gamma] &=&
\frac{256 \pi^3\alpha^3 \nu^2\, |\gamma_X|\, M_{*0}^2 M_X^3}{3} \left( \frac{s-M_X^2}{s} \right)^5
 \left| F(s,M_X^2 \!-\!i M_X\Gamma_X) \right|^2
\nonumber\\
&& \hspace{4cm}
\times \left[ \left| A_0 - \frac{1}{\sqrt{5}} A_2\right|^2 + \frac{9}{20}|A_2|^2 \right] .
\label{sigmagammaXthresh}
\end{eqnarray}

The factor in square brackets in Eq.~\eqref{sigmagammaXthresh} that depends on $A_0$  and $A_2$ 
differs from the factor $|A_0|^2  + |A_2|^2$ in the cross section for producing $D^{*0} \bar D^{*0}$ 
in Eq.~\eqref{sigma-threshold} by a multiplicative factor that depends on the complex ratio $A_2/A_0$.
If $|A_0|^2  + |A_2|^2$ has been determined  but $|A_2|/|A_0|$ has not,
we should allow for all possible complex values of $A_2/A_0$.
For the value of $|A_0|^2  + |A_2|^2$
 determined by Eq.~\eqref{A0,A2}, the multiplicative factor can range from 0.34 to 1.31. 
If  $|A_0|^2$  and $ |A_2|^2$ have both been determined,
we must still allow for all possible phases of $A_2/A_0$.
For the values of $|A_0|$ and $|A_2|$ in Eq.~\eqref{A0,A2},
the multiplicative factor can range from 0.36 to 1.10,
and it is equal to 0.73  if the phase of $A_2/A_0$ is $\pm i$.

\begin{figure}[hbt]
\includegraphics*[width=0.7\linewidth]{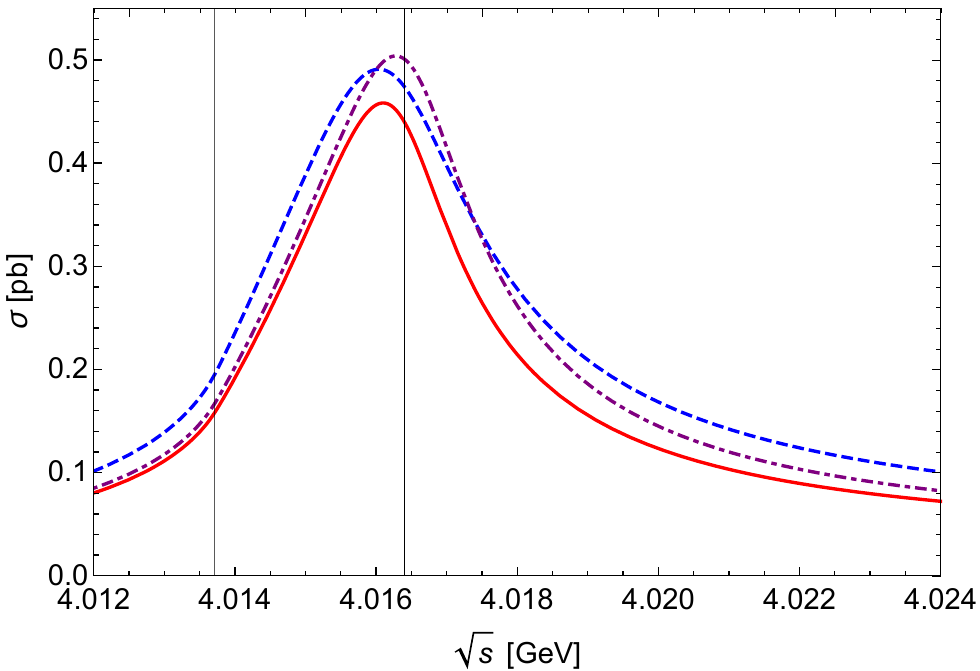} 
\caption{
Cross section for producing $X+\gamma$ as a function of the center-of-mass energy $\sqrt{s}$.  
The binding energy and width of $X$ are 
$|E_X| = 0.05$~MeV, $\Gamma_X = \Gamma_{*0}$ (solid red curve),
$|E_X| = 0.10$~MeV, $\Gamma_X = \Gamma_{*0}$ (dashed blue curve), 
and $|E_X| = 0.05$~MeV, $\Gamma_X = 2\, \Gamma_{*0}$  (dot-dashed purple curve).
The two vertical lines are at the $D^{*0} \bar D^{*0}$ threshold and at the 
triangle-singularity energy $\sqrt{s_\triangle}$.
 }
\label{fig:sigmaXgamma}
\end{figure}

The cross section for $X+\gamma$ predicted using  Eq.~\eqref{sigmagammaXthresh} 
 is shown in Fig.~\ref{fig:sigmaXgamma}
for center-of-mass energies  up to about 10~MeV above the $D^{*0} \bar D^{*0}$ threshold.
The absolute values of the amplitudes $A_0$ and $A_2$ are given in Eq.~\eqref{A0,A2}.
For purposes of illustration, we choose the complex phase of $A_2/A_0$ to be $\pm i$.
The cross section is shown for the first three choices of binding energy and width of $X$
enumerated after Eq.~\eqref{EX,GammaX}.
The cross section has a narrow  peak near the  energy for the triangle singularity predicted by Eq.~\eqref{s-triangle}.
For $|E_X| = 0.05$~MeV, $\Gamma_X = \Gamma_{*0}$, the peak in the cross section is at 4016.1~MeV, 
which is only 0.3~MeV lower than the prediction from Eq.~\eqref{s-triangle}.
Doubling the binding energy $|E_X|$ has very little effect on the position of the peak in the cross section,
but it increases its height by about 7\%.
Doubling the width $\Gamma_X$ moves the peak  to an energy higher by about 0.2~MeV
and increases its height by about 10\%.
The peak in the cross section in Fig.~\ref{fig:sigmaXgamma} is large enough that it may be observable by the BESIII collaboration.
Its energy is in a range not covered by previous measurements of the cross section
for $e^+ e^- \to X+\gamma$ by the BESIII collaboration \cite{Ablikim:2013dyn,Ablikim:2019zio}.

In Ref.~\cite{Braaten:2019gwc}, the cross section for $e^+e^-\to X +\gamma$ was calculated using 
nonrelativistic approximations.  The result is consistent with that in Eq.~\eqref{sigmagammaXthresh} 
to within the accuracy of the approximations.  
The prefactors differ by terms suppressed by $(\delta/M_{*0})^2$.
The nonrelativistic approximation to the loop amplitude $F(s,u)$  is a function $F(W,U)$ 
of the center-of-mass energy $W=\sqrt{s} - 2 M_{*0}$ relative to the $D^{*0} \bar D^{*0}$ threshold in the $e^+ e^-$ CM frame
and the energy $U = \sqrt{u} -(M_{*0} + M_0)$
of the $D^{*0} \bar{D}^0$ pair relative to their threshold in their CM frame.
The function $F(W,U)$ is given analytically in Eq.~\eqref{FanalyticNR} of Appendix~\ref{LoopIntegralNR}.
In Ref.~\cite{Braaten:2019gwc}, that function was evaluated at the negative energy $U = E_X$.
It should more appropriately be evaluated at the complex energy $U = E_X-i\Gamma_X/2$.
The difference has a negligible effect on the numerical results in  Ref.~\cite{Braaten:2019gwc}.

\begin{figure}[hbt]
\includegraphics*[width=0.7\linewidth]{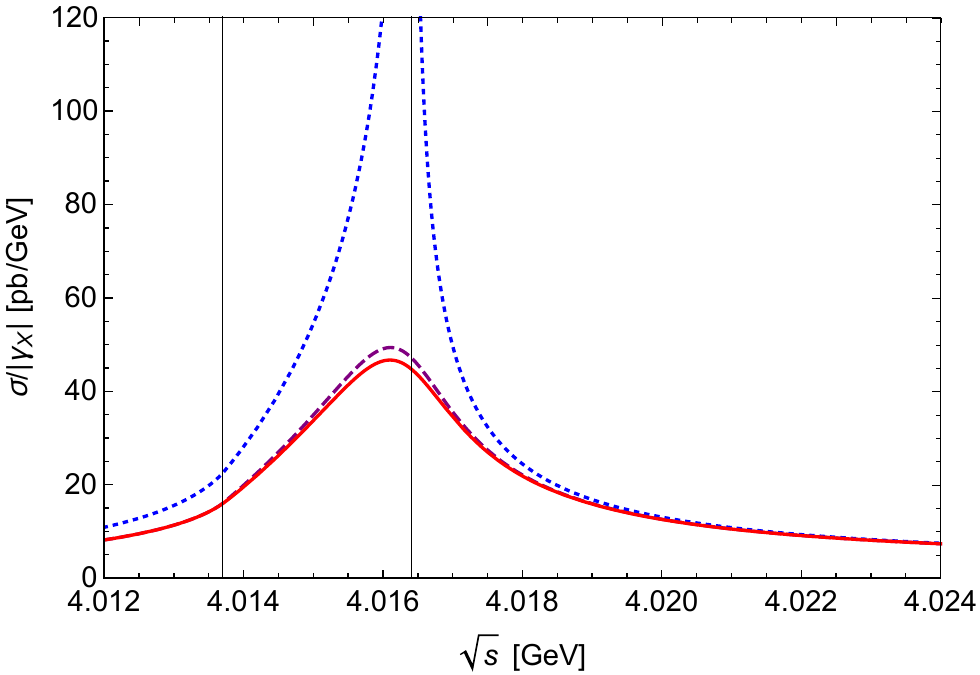} 
\caption{
Cross section for producing $X+\gamma$ divided by $|\gamma_X|$
as a function of the center-of-mass energy $\sqrt{s}$.  
The binding energy and width of $X$ are 
$|E_X| = 0.05$~MeV, $\Gamma_X = \Gamma_{*0}$ (solid red curve),
$|E_X| = 0.05$~MeV, $\Gamma_X = 0$ (dashed purple curve),
and $|E_X| = 0$, $\Gamma_X = 0$ (dotted blue curve).
The two vertical lines are at the $D^{*0} \bar D^{*0}$ threshold and at the 
triangle-singularity energy $\sqrt{s_\triangle}$.
 }
\label{fig:sigmaXgammatri}
\end{figure}

The origin of the peak in the cross section in Fig.~\ref{fig:sigmaXgamma} is the logarithmic branch point 
of the loop amplitude $F(s,M_X^2 \!-\!i M_X\Gamma_X)$ in Eq.~\eqref{sigmagammaXthresh}
at a complex value of $s$ near $s_\triangle$  in Eq.~\eqref{s-triangle}.
An analytic approximation to the logarithmic term in $F(s,u)$  can be obtained by setting
$W=\sqrt{s} - 2 M_{*0}$ and $U = E_X \!-\! i \Gamma_X/2$
in the nonrelativistic loop amplitude $F(W,U)$  in Eq.~\eqref{FanalyticNR}.
In the limit as the $D^{*0}$ width $\Gamma_{*0}$, the binding energy $E_X$,
and the width $\Gamma_X$ all go to 0,
the branch point approaches the real $s$ axis and the loop amplitude has a 
logarithmic divergence proportional to $\log|s-s_\triangle|$.
However the cross section goes to 0 in this limit 
because of the factor of $|\gamma_X|$ in Eq.~\eqref{sigmagammaXthresh}.
The divergence from the triangle singularity is
illustrated in Fig.~\ref{fig:sigmaXgammatri}, which shows $\sigma/|\gamma_X|$ as a function of $\sqrt{s}$
for the values of $E_X$, $\Gamma_X$ numbered 1, 4, and 5 after Eq.~\eqref{EX,GammaX}.
The curve for $|E_X| = 0$, $\Gamma_X = 0$ 
has a $\log^2(|s -s_\triangle|)$ divergence.
The curve for $|E_X| = 0.05$~MeV, $\Gamma_X = 0$ shows that  the divergence is 
reduced to a narrow peak by the suppression from the binding energy only.
The curve for $|E_X| = 0.05$~MeV, $\Gamma_X = \Gamma_{*0}$ 
shows that the $D^{*0}$ width provides some small additional suppression of the peak.


\section{Production of   $\bm{D^{*0} \bar{D}^0 + \gamma}$  near  the $\bm{D^{*0} \bar{D}^{*0}}$ threshold}
\label{sec:DDgamma}

\begin{figure}[hbt]
\includegraphics*[width=0.35\linewidth]{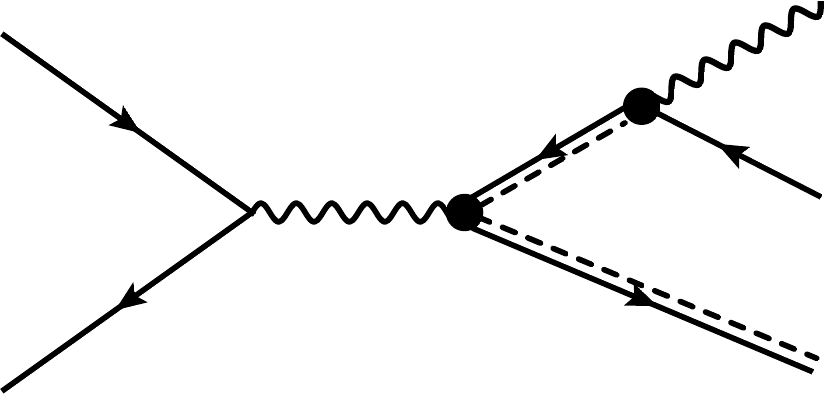}
\caption{Tree diagram for $e^+e^-\to D^{*0} \bar{D}^0 +\gamma$.
}
\label{fig:eetogammaDDtree}
\end{figure}

\begin{figure}[hbt]
\includegraphics*[width=0.35\linewidth]{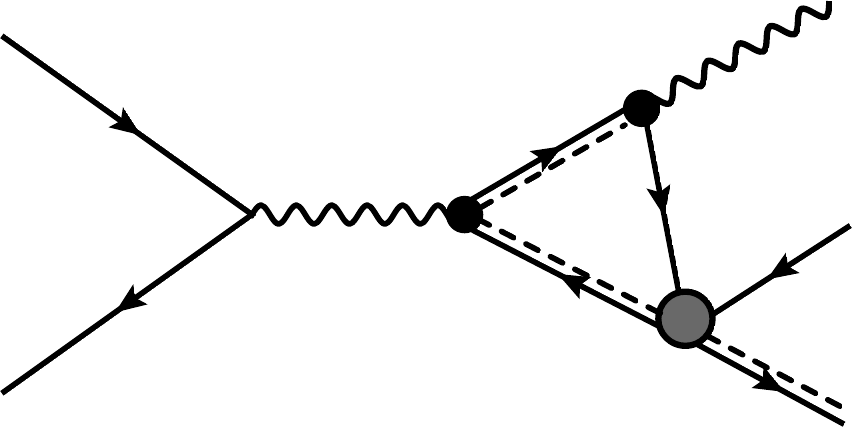} \quad
\includegraphics*[width=0.35\linewidth]{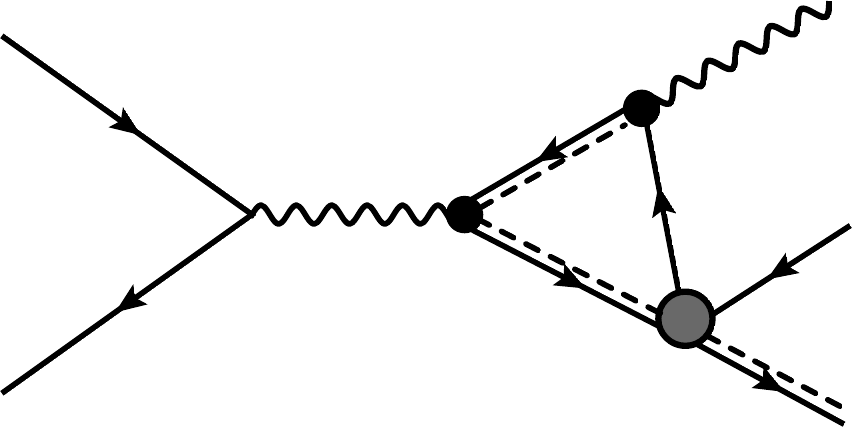} 
\caption{Feynman diagrams for $e^+e^-\to D^{*0} \bar{D}^0 +\gamma$ with a charm-meson loop.
The grey blob represents the charm-meson elastic scattering amplitude.
}
\label{fig:eetogammaDDloop}
\end{figure}

Since $e^+e^-$ annihilation can produce $X(3872)$  recoiling against  a photon,
it can also produce its constituents $D^{*0} \bar D^0$ recoiling against  a photon.
The virtual photon from $e^+e^-$ annihilation creates the charm mesons $D^{*0}$ and $ \bar D^{*0}$,
and the real photon is produced by a subsequent radiative transition $\bar D^{*0} \to \bar D^0$.
The final state $D^{*0} \bar D^0 +\gamma$ can either
be produced directly through the tree diagram in Fig.~\ref{fig:eetogammaDDtree}
or  through the loop diagrams in Fig.~\ref{fig:eetogammaDDloop}.
The tree diagram is enhanced by the $\bar D^{*0}$ resonance.
The loop diagrams have a charm-meson triangle.

The loop diagrams in Fig.~\ref{fig:eetogammaDDloop} involve the amplitudes for the scattering 
of $D^{*0} \bar D^0$ or $D^0 \bar D^{*0}$ into $D^{*0} \bar D^0$.
A Lorentz-covariant expression for the vertex for either of those amplitudes is
\begin{equation}
\frac{-i (4M_{*0} M_0)\pi/\mu}
{-\gamma_X - i \, \lambda^{1/2}(P^2,M_{*0}^2 \!-\! i M_{*0} \Gamma_{*0}, M_0^2)/\big(2 \sqrt{P^2}\, \big)}
\, (g^{\mu\nu} - P^\mu P^\nu/P^2) ,
\label{DDtoDD}
\end{equation}
where $P$ is the total momentum of the charm-meson pair,
$\mu$ and $\nu$ are the Lorentz indices of the spin-1 charm mesons,
$\gamma_X$ is the $X$ binding momentum,  and
$\lambda(x,y,z)$ is the K\"allen function defined after Eq.~\eqref{Gamma*ratio}.
If we set $\Gamma_{*0} = 0$, the second term in the denominator 
reduces to $-i k_\mathrm{rel}$, where $k_\mathrm{rel}$ is the relative 3-momentum 
 of $D^{*0} \bar D^0$ in their center-of-momentum frame.
In that frame, the vertex in Eq.~\eqref{DDtoDD} reduces to the nonrelativistic scattering amplitude 
$i(\pi/\mu)\delta^{mn}/(-\gamma_X - i k_\mathrm{rel})$
multiplied by a factor $4M_{*0} M_0$ that compensates for the relativistic normalization of states.
Aside from this normalization factor, the vertex  depends on the mass and decay width of the $D^{*0}$ 
 only through the argument $M_{*0}^2 - i M_{*0} \Gamma_{*0}$  of the K\"allen function.  
 The vertex in Eq.~\eqref{DDtoDD} is a good approximation only if $u=P^2$ is close enough to the
$D^{*0}\bar D^0$ threshold $M_X^2$.   It should remain a good approximation 
up to about the  $D^{*+}D^-$ threshold, which is higher than $M_X$ by 8.2~MeV.

\begin{figure}[hbt]
\includegraphics*[width=0.7\linewidth]{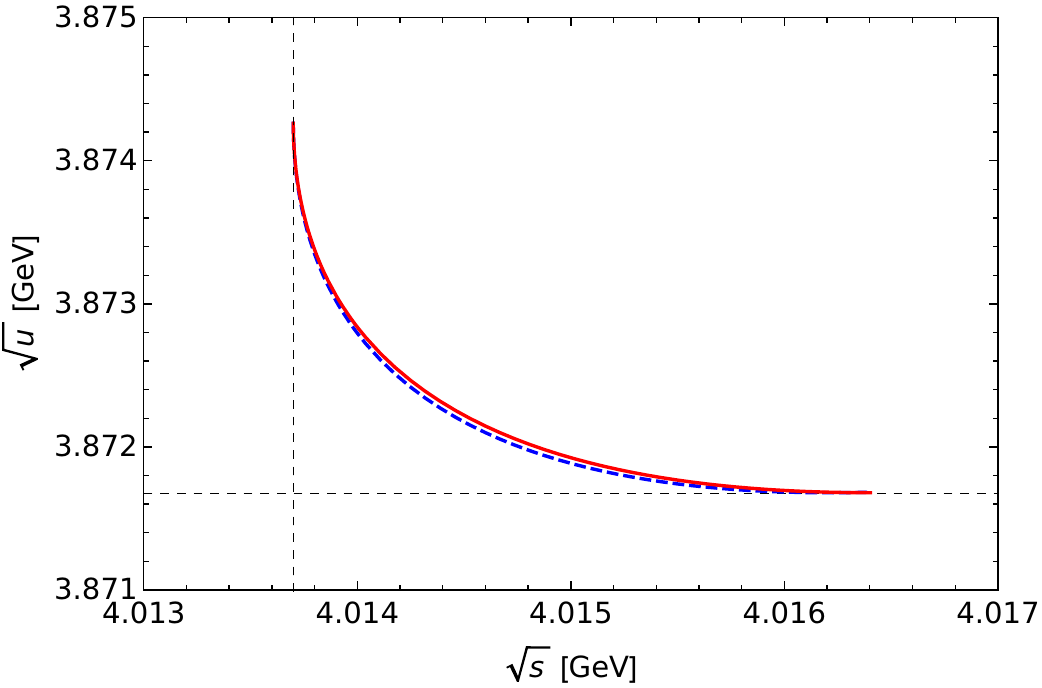}
\caption{
The  charm-meson-pair invariant mass $\sqrt{u_\triangle}$ for
the triangle singularity as a function of the center-of-mass energy $\sqrt{s}$:
the exact result in Eq.~\eqref{u-triangle} (solid red curve)
and the approximation in Eq.~\eqref{u-triangle:expand} (dashed blue curve).
The vertical and horizontal dashed lines are $\sqrt{s} =  2M_{*0}$ and $\sqrt{u} =M_X$, respectively.
}
\label{fig:triangle-us}
\end{figure}

The loop integrals for the diagrams in Fig.~\ref{fig:eetogammaDDloop} produce a triangle singularity
from the region where the three charm mesons that form the loop are all on their mass shells.
The triangle singularity produces a logarithmic branch point in $F(s,u)$ 
along a line of complex values of $s$ near $4 M_{*0}^2$ and  $u$ near $M_X^2$.
In the limit $\Gamma_{*0} \to 0$, the line approaches real values of $s$ and $u$.
In Appendix~\ref{LoopIntegralR}, the condition for the triangle singularity is deduced from the on-shell conditions 
for the three charm mesons in the triangle.
The range of $s$ for the triangle singularity is $4M_{*0}^2 < s \le s_\triangle$,
where $s_\triangle$ is given in Eq.~\eqref{s-triangle}.
The solution for $u$ as a function of $s$ is
\begin{equation}
u_\triangle(s) =(M_{*0} + M_0)^2
+  \frac{ \big[ (M_{*0} - M_0)\sqrt{s} - (M_{*0} + M_0)\sqrt{s-4M_{*0}^2} \big]^2}{ 4 M_{*0}^2} .
\label{u-triangle}
\end{equation}
The function $u_\triangle(s)$ is illustrated in Fig.~\ref{fig:triangle-us}.
As $s$ increases from $4M_{*0}^2$ to $s_\triangle$, $u_\triangle$ decreases 
from $2(M_{*0}^2+M_0^2)$ to $(M_{*0} + M_0)^2$.
Numerically, as $\sqrt{s}$ increases from 4013.7~MeV to 4016.4~MeV,
$u_\triangle$ decreases from 15.010~GeV$^2$ to 14.990~GeV$^2$.
The triangle singularity appears only in narrow intervals of $s$ and $u$ 
with lengths $(M_{*0}/M_0)\delta^2$ and $\delta^2$, respectively.
If we set $M_0=M_{*0} - \delta$ and take $s - 4 M_{*0}^2$ to be order $\delta^2$,  
the solution for $u$ at the triangle singularity 
in Eq.~\eqref{u-triangle} can be expanded in powers of $\delta$.
The expansion to order $\delta^2$ can be expressed as
\begin{equation}
u_\triangle(s) \approx M_X^2 +  \left( \delta - \sqrt{s-4M_{*0}^2}\,  \right)^2.
\label{u-triangle:expand}
\end{equation}
In  Fig.~\ref{fig:triangle-us},  this approximation is compared to the exact result in Eq.~\eqref{u-triangle}.
The expansion to order $\delta^3$ would be difficult to distinguish from the exact result.

We consider the production of $D^{*0}$, $\bar D^0$, and $\gamma$ 
with 4-momenta $p_1$, $p_0$, and $q$ and with  total 4-momentum $Q=p_1+p_0+q$.
A convenient set of Lorentz scalars is the center-of-mass energy  squared $s=Q^2$ and
the squares $u=(p_1+p_0)^2$  and $t=(p_0+q)^2$ of the invariant masses 
of $D^{*0} \bar D^0$ and $\bar D^0 \gamma$, respectively.
The matrix element from the tree diagram  in Fig.~\ref{fig:eetogammaDDtree}  is
\begin{equation}
\mathcal{M}_\mathrm{tree} = 
 \frac{2 e^3\nu \sqrt{4M_{*0} M_0}}{s\, (t - M_{*0}^2 + i M_{*0} \Gamma_{*0})}
\bar v \gamma_\mu u\, \mathcal{A}(Q)^{\mu\nu\lambda\sigma} p_{1\nu} \epsilon_{\sigma \alpha \beta \tau }\,
  p_0^\alpha q^\beta\, \varepsilon^*(p_1)_\lambda \, \varepsilon^*(q)^\tau.
\label{MDDgamma-tree}
\end{equation}
The loop integral from the loop diagrams  in Fig.~\ref{fig:eetogammaDDloop} 
can be simplified in the region near the triangle singularity
where $s-4M_{*0}^2$ and $u-M_X^2$ are both order $\delta^2$.
In Appendix~\ref{LoopIntegralR}, the leading term in an expansion in powers of $\delta/M_{*0}$
is reduced to the function  $F(s,u)$ 
defined by the Lorentz-invariant momentum integral in Eq.~\eqref{FLorentz}.
The charm-meson loop can be reduced to an effective vertex for the coupling of the virtual photon 
to $D^{*0}$, $ \bar D^0$, and a real photon that is given in Eq.~\eqref{effvertex:D*Dgamma}.
The  matrix element from the loop diagrams  in Fig.~\ref{fig:eetogammaDDloop}  is
\begin{eqnarray}
\mathcal{M}_\mathrm{loop}  &=& 
\frac{8 \pi  e^3\nu M_{X}  }{s}\,
 \frac{\sqrt{4M_{*0} M_0} }
{-\gamma_X - i \, \lambda^{1/2}(u,M_{*0}^2 \!-\! i M_{*0} \Gamma_{*0}, M_0^2)/\big( 2\sqrt{u}\, \big)} 
\,F(s,u)
\nonumber\\
&& \hspace{2cm}\times \,
\bar v \gamma_\mu u  \,
\mathcal{A}(Q)^{\mu \nu \lambda \sigma} q_\nu \, \left( g_{\lambda\rho} - \frac{P_\lambda P_\rho}{u} \right)
\epsilon_{\sigma\alpha \beta \tau}Q^\alpha q^\beta\, \varepsilon^*(p_1)^\rho \, \varepsilon^*(q)^\tau  ,
\label{MDDgamma-loop}
\end{eqnarray}
where $P = p_0 + p_1$.
Our matrix elements $\mathcal{M}_\mathrm{tree}$ in Eq.~\eqref{MDDgamma-tree}
and  $\mathcal{M}_\mathrm{loop}$  in Eq.~\eqref{MDDgamma-loop} are good approximations 
only if $\sqrt{s}$ is less than about 10~MeV above the $D^{*0}  \bar D^{*0}$ threshold at  4013.7~MeV,
because this is the region of validity of our expression for $\mathcal{A}(Q)^{\mu \nu \lambda \sigma} $
in Eq.~\eqref{A[eetoD*D*]} with constant amplitudes $A_0$ and $A_2$.
Our expression for  $\mathcal{M}_\mathrm{loop}$ should be a good approximation  for  $\sqrt{u}$
up to about the $D^{*+} D^-$ threshold at 3879.9~MeV,
because this is the region of validity of the elastic scattering amplitude for the charm-meson pair in Eq.~\eqref{DDtoDD}.

At a given center-of-mass energy $\sqrt{s}$, the square of the matrix element summed over spins 
is  a  function of $u$, $t$, and two angles.
The differential cross section integrated over angles and summed over final spins is
\begin{equation}
\frac{d \sigma}{du\, dt}= \frac{1}{256 \pi^3\, s^2} 
\left\langle \big|\mathcal{M}_\mathrm{tree} + \mathcal{M}_\mathrm{loop} \big|^2 \right\rangle.
\label{dsigma/dudt}
\end{equation}
The angular brackets indicate the average over angles.
The matrix elements in Eqs.~\eqref{MDDgamma-tree} and \eqref{MDDgamma-loop} both depend 
on the  amplitudes $A_0$ and $A_2$ in the factor $\mathcal{A}(Q)^{\mu \nu \lambda \sigma}$.
Their absolute values are given in Eq.~\eqref{A0,A2}.
For purposes of illustration, we choose the complex phase of $A_2/A_0$ to be $\pm i$.
The matrix element $\mathcal{M}_\mathrm{loop}$ also depends on the binding momentum
$\gamma_X$ through the denominator  in Eq.~\eqref{MDDgamma-loop}.

The scatter plot for the differential cross section in the $u-t$ plane is called a {\it Dalitz plot}.
The boundary of the Dalitz plot is the curve defined by the equation
\begin{eqnarray}
0 &=& t\, u^2 + t^2\, u - (s+M_{*0}^2+M_0^2)\, t\,u
\nonumber\\
&&-M_0^2(s-M_{*0}^2)\, u +(M_{*0}^2-M_0^2)s\, t + M_0^2s(s-M_{*0}^2+M_0^2) .
\label{Dalitz-boundary}
\end{eqnarray}
This function of $u$ and $t$ is negative inside the boundaries of the Dalitz plot.
The values of $u$ range from $(M_{*0}+M_0)^2$ to $s$.
The values of $t$ range from $M_0^2$ to  $( \sqrt{s}\ - M_{*0} )^2$.
Our matrix elements $\mathcal{M}_\mathrm{tree}$ in Eq.~\eqref{MDDgamma-tree}
and  $\mathcal{M}_\mathrm{loop}$  in Eq.~\eqref{MDDgamma-loop} are good approximations for
$\sqrt{s}$ less than about  4.024~GeV and in the corner of the Dalitz plot with
$u$ less than about 15.054~GeV$^2$.

\begin{figure}[hbt]
\includegraphics*[width=0.7\linewidth]{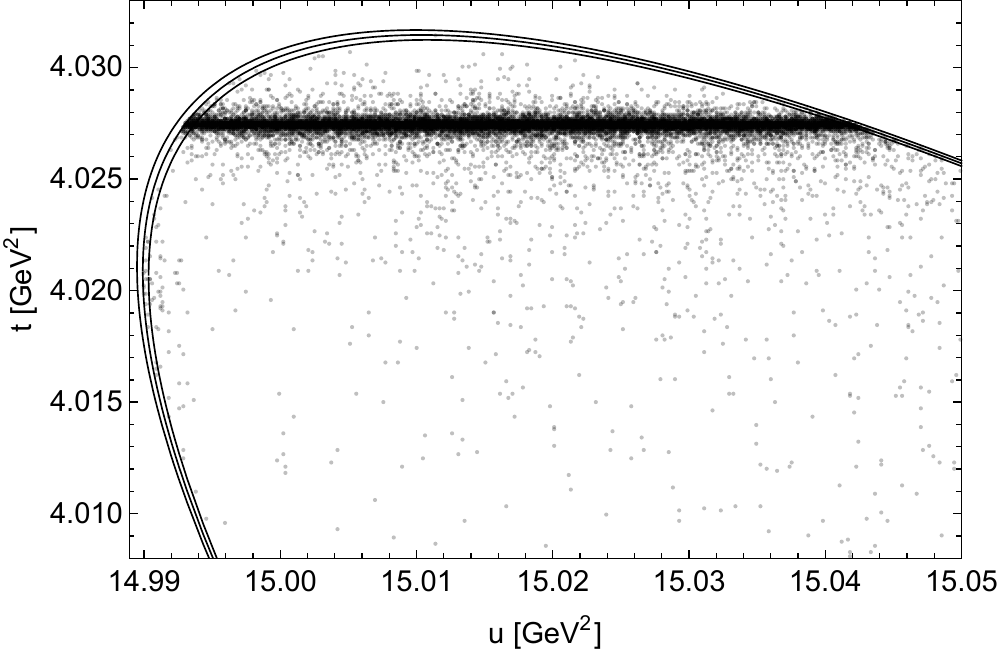} 
\caption{
Upper left corner of the Dalitz plot for $D^{*0} \bar D^0 + \gamma$
at $\sqrt{s}=4014.7$~MeV for $\Gamma_{*0}=55$~keV.
The binding energy and width of $X$ are $|E_X| = 0.05$~MeV, $\Gamma_X = \Gamma_{*0}$.
The three curves are the boundaries of the Dalitz plot  if the $D^{*0}$ mass is set equal to
$M_{*0} - \Gamma _{*0}$, $M_{*0}$, and $M_{*0} + \Gamma _{*0}$.
The horizontal band is from the $\bar D^{*0}$ resonance.
The triangle-singularity line is predicted to be at $u_\triangle=14.993~\mathrm{GeV}^2$.
This vertical line and the horizontal line at the center of the resonance band
intersect on the boundary of the Dalitz plot. 
There are 20,000 events in the Dalitz plot, 97 of which are to the left of $u_\triangle$.
}
\label{fig:Dalitzcorner}
\end{figure}

To illustrate the behavior of the  differential cross section as a function of $u$ and $t$,
 we choose the center-of-mass energy  $\sqrt{s} = 4014.7$~MeV, which is 1~MeV
above the $D^{*0} \bar{D}^{*0} $ threshold and near the middle of the triangle-singularity region.
The triangle-singularity line $u=u_\triangle(s)$ has two endpoints on the boundary of the Dalitz plot.
The upper endpoint coincides with the point where the horizontal line $t=M_{*0}^2$ 
at the center of the $\bar{D}^{*0} $ resonance band cuts through the boundary of the Dalitz plot.
At this point, the $D^{*0}$ and $\bar D^0$ are moving in the same direction in the center-of-momentum frame.
A blow-up of the upper left corner of the Dalitz plot for $\sqrt{s} = 4014.7$~MeV  is shown in Fig.~\ref{fig:Dalitzcorner}.
The boundary of the Dalitz plot is actually a little fuzzy on the scale in Fig.~\ref{fig:Dalitzcorner}, 
because the $D^{*0}$ has a nonzero width.  This effect is illustrated by showing three boundaries 
corresponding to setting the $D^{*0}$ mass equal to
$M_{*0} - \Gamma _{*0}$, $M_{*0}$, and $M_{*0} + \Gamma _{*0}$.

To generate events populating the Dalitz plot, 
we first generate random dots  in the region of the $u-t$ plane shown in Fig.~\ref{fig:Dalitzcorner}.
For each point, we generate a random number between 0 and 1.
We keep the dot in the plot if  $d\sigma/dudt$ at that point 
is larger than the product of the random number and the maximum value of $d\sigma/dudt$ in the region.
The probability of keeping the dot is the ratio of $d\sigma/dudt$ to its maximum value.
The density of dots is therefore proportional to the differential cross section.
The ratio of the number of dots kept to the number of dots generated is about 1\%.
The resulting Dalitz plot at a center-of-mass energy 1~MeV above the $D^{*0} \bar{D}^{*0} $ threshold
is shown in Fig.~\ref{fig:Dalitzcorner}.
There is an obvious horizontal band  near $t=M_{*0}^2$ from the $\bar{D}^{*0} $ resonance.
The triangle singularity would be expected to produce a narrow vertical band in the Dalitz plot near the 
value $u_\triangle =14.993~\mathrm{GeV}^2$ predicted by Eq.~\eqref{u-triangle}.  
There is no narrow vertical band near that value of $u$.
However  the density of points to the left of $u_\triangle$
is significantly larger than that just to the right of $u_\triangle$.

\begin{figure}[hbt]
\includegraphics*[width=0.7\linewidth]{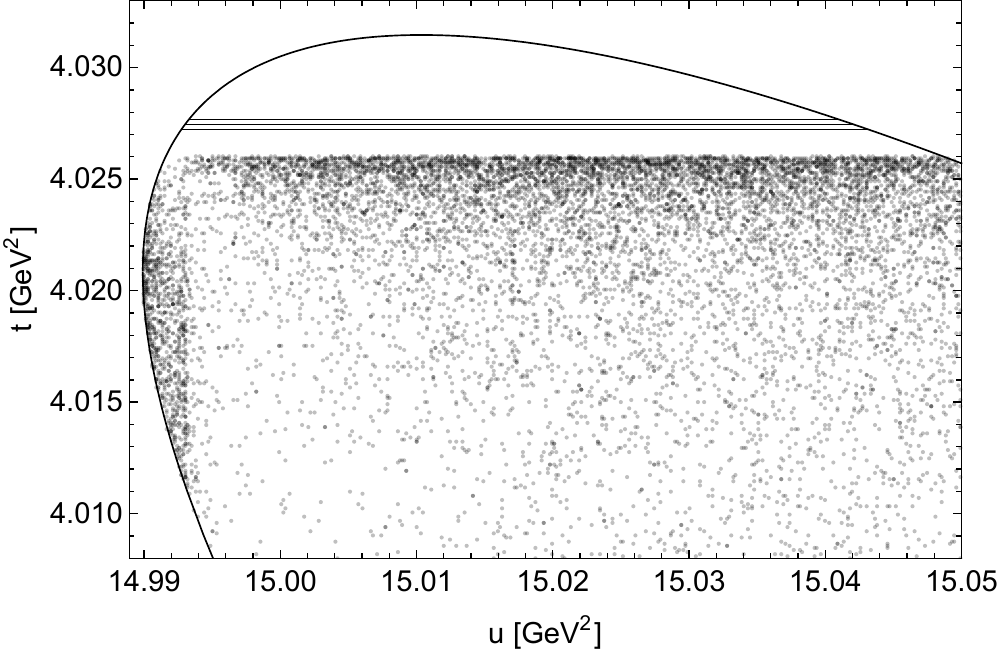} 
\caption{
Upper left corner of the Dalitz plot for $D^{*0} \bar D^0 + \gamma$
at $\sqrt{s}=4014.7$~MeV for $\Gamma_{*0}=0$.
The binding energy and width of $X$ are $|E_X| = 0.05$~MeV, $\Gamma_X = 0$.
The horizontal lines are at three values of the $\bar D^{*0} \gamma$ invariant mass-squared $t$:
$(M_{*0}+\Gamma_{*0})^2$, $M_{*0}^2$, and $(M_{*0}-\Gamma_{*0})^2$ for $\Gamma_{*0} = 55$~keV.
The differential cross section diverges at $t=M_{*0}^2$,
so events were generated only in the region $t<4.026~\mathrm{GeV}^2$.
There is a narrow vertical band along
the triangle-singularity line  at $u_\triangle=14.993~\mathrm{GeV}^2$.
There are 10,000 events in the Dalitz plot, 865 of which are to the left of $u_\triangle$.
}
\label{fig:Dalitzcorner0}
\end{figure}

The effects of the triangle singularity on the Dalitz plot would be more distinct if the $D^{*0}$ width
$\Gamma_{*0}$ were smaller.
The $\bar D^{*0}$ resonance band would be narrower in proportion to $\Gamma_{*0}$,
and the number of events in the resonance band would be larger in proportion to $1/\Gamma_{*0}$.
The Dalitz plot at a center-of-mass energy 1~MeV above the $D^{*0} \bar{D}^{*0} $ threshold
is shown in Fig.~\ref{fig:Dalitzcorner0} for the limiting case $\Gamma_{*0} \to 0$.
Since $d\sigma/dudt$ diverges at $t=M_{*0}^2$, 
events were generated only in the region $t<4.026~\mathrm{GeV}^2$.
A faint narrow vertical band is visible along the triangle-singularity line 
at $u_\triangle =14.993~\mathrm{GeV}^2$ predicted by Eq.~\eqref{u-triangle}.  
A more obvious feature is that the density of points to the left of $u_\triangle$
is much larger than that just to the right of $u_\triangle$.

\begin{figure}[hbt]
\includegraphics*[width=0.7\linewidth]{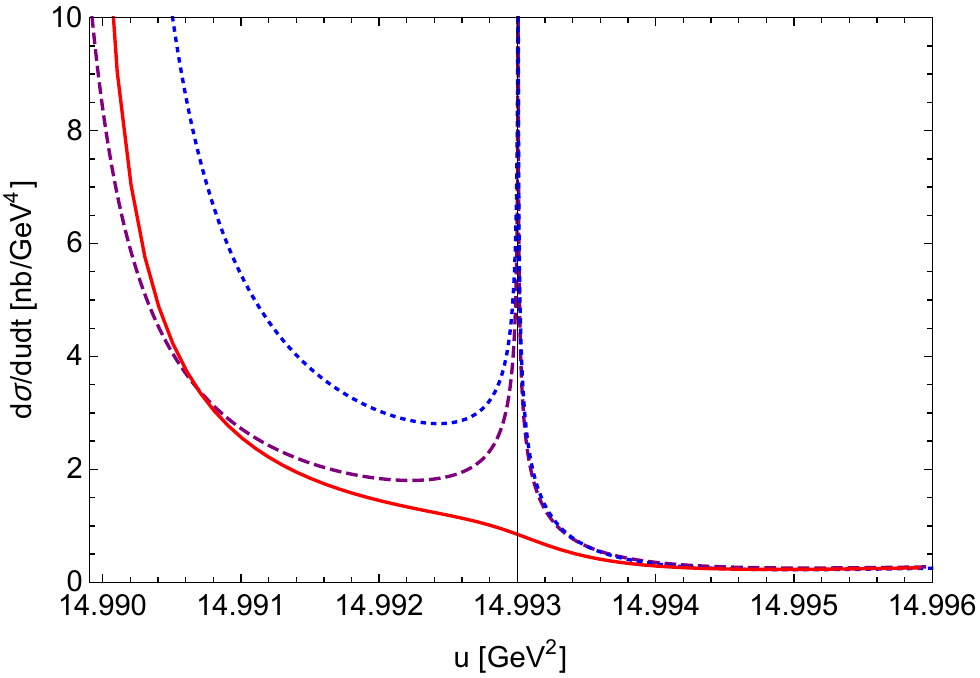} 
\caption{
Differential cross section $d\sigma/dudt$  for producing $D^{*0} \bar D^0 + \gamma$
as a function of $u$ at  $\sqrt{s}=4014.7$~MeV and $t=4.0209~\mathrm{GeV}^2$. 
The binding energy and width of $X$ are 
$|E_X| = 0.05$~MeV, $\Gamma_X = \Gamma_{*0}$ (solid red curve),
$|E_X| = 0.05$~MeV, $\Gamma_X = 0$  (dashed purple curve),
 and $|E_X| = 0$, $\Gamma_X =0$ (dotted  blue curve).
The vertical line is at the predicted value $u_\triangle=14.993~\mathrm{GeV}^2$ for the triangle singularity.
 }
\label{fig:dsigma/dudt}
\end{figure}

In Fig.~\ref{fig:dsigma/dudt}, we illustrate  the differential cross section $d\sigma/dudt$
as a function of $u$ in the region near the predicted triangle singularity at $u_\triangle$ for $\sqrt{s} = 4014.7$~MeV. 
We choose the $\bar{D}^{*0} \gamma$ invariant mass-squared to be 
its value at the left-most point of the Dalitz plot in Fig.~\ref{fig:Dalitzcorner}:
$t=4.0209~\mathrm{GeV}^2$. 
We compare $d\sigma/dudt$ for the values of $E_X$, $\Gamma_X$ numbered 1, 4, and 5 
after Eq.~\eqref{EX,GammaX}.
For $E_X=0$, $\Gamma_{*0} = 0$, $d\sigma/dudt$ diverges at $u=u_\triangle$, with a shape consistent 
with a log$^2$ divergence.
For $|E_X| = 0.05$~MeV, $\Gamma_X = 0$,
$d\sigma/dudt$ also diverges at $u=u_\triangle$ but the peak is a little narrower.
For $|E_X| = 0.05$~MeV, $\Gamma_X=\Gamma_{*0}$,
$d\sigma/dudt$ decreases monotonically with $u$ out to beyond the triangle singularity.
The suppression from the physical $D^{*0}$ width is large enough that there is not even
a local maximum near $u_\triangle$.


\section{Schmid Cancellation}
\label{sec:Schmid}

The effects of the triangle singularity on the cross section for $e^+e^-$ annihilation into $D^{*0} \bar D^0 + \gamma$
are not obvious in the Dalitz plot in Fig.~\ref{fig:Dalitzcorner} or in the differential cross section in 
Fig.~\ref{fig:dsigma/dudt} for the physical value of the $D^{*0}$ width.
The effects of a triangle singularity can be suppressed by nonzero widths of particles in
the triangle.  They can also be suppressed by cancellations between diagrams.
In the limit $\Gamma_{*0} \to 0$,
the triangle singularity in $e^+ e^-$ annihilation into $D^{*0} \bar D^0 + \gamma$
produces a log$^2$ divergence in the differential cross section $d\sigma/dudt$
along the  triangle-singularity line $u= u_\triangle(s)$ with $t$ below the $\bar D^{*0}$ resonance.
Schmid pointed out that in the differential cross section $d\sigma/du$ integrated over $t$, 
the triangle singularity is suppressed by interference
between the loop diagrams in Fig.~\ref{fig:eetogammaDDloop}
and  the tree diagram in Fig.~\ref{fig:eetogammaDDtree} \cite{Schmid:1967ojm}.
The  cancellation is between the loop contribution along the triangle-singularity line
in the region of $t$ below the resonance band 
and the interference contribution from the region of $t$ in the resonance band. 
Anisovich and Anisovich pointed out that the divergence in $d\sigma/du$ is not completely cancelled,
but the log$^2$ divergence is reduced to a single logarithm  \cite{Anisovich:1995ab}.
Dalitz plot distributions in the presence of triangle singularities have been studied in detail by
Szczepaniak \cite{Szczepaniak:2015hya}.
The effects on the Schmid cancellation from the decay widths of particles in the triangle loop
have been considered by Debastiani, Sakai and Oset \cite{Debastiani:2018xoi}.

The differential cross section   for $e^+ e^-$ annihilation into $D^{*0} \bar D^0 + \gamma$
averaged over angles is given in Eq.~\eqref{dsigma/dudt}.
The matrix elements $\mathcal{M}_\mathrm{tree}$ and $ \mathcal{M}_\mathrm{loop}$
are given in Eqs.~\eqref{MDDgamma-tree} and \eqref{MDDgamma-loop}.
The full expression for  $d\sigma/du\,dt$  is very complicated.
To exhibit the Schmid cancellation, it is helpful to simplify the cross section
in the region of $s$, $u$, and $t$ near the triangle singularity.
If $s-4M_{*0}^2$ and  $u-M_X^2$ are both order $\delta^2$,
the kinematic constraints on the Dalitz plot require $t-M_{*0}^2$ to  also be order $\delta^2$. 
It is convenient to introduce subtracted variables:
\begin{equation}
s_- = s-4M_{*0}^2, \qquad u_-=u-M_X^2, \qquad t_-=t-M_{*0}^2.
\label{Dalitz-variables-small}
\end{equation}
In the region where $s_-$, $u_-$, and $t_-$ are all order $\delta^2$,
the equation for the boundary of the Dalitz plot in Eq.~\eqref{Dalitz-boundary} reduces
  to $0= D(s_-,u_-,t_-)$, where
\begin{equation}
D(s_-,u_-,t_-) = -\delta^4 + 2 \delta^2 (s_- \!+\! u_-  \!-\! 2t_-)- (s_-  \!-\! u_- \!-\! 2t_-)^2 .
\label{Dalitz-boundary-small}
\end{equation}
The function $D(s_-,u_-,t_-)$ is positive inside the boundaries of the Dalitz plot.
The triangle singularity condition in Eq.~\eqref{u-triangle:expand} reduces at leading order in $\delta$ to
$\sqrt{u_-} + \sqrt{s_-}= \delta$ with $0 < s_- < \delta^2$.

To simplify the differential cross section, we set $M_0 = M_{*0} - \delta$ and
we take $s_-$, $u_-$, and $t_-$ to be order $\delta^2$.
We also take $\gamma_X$ to be order $\delta$ and $M_{*0}\Gamma_{*0}$  to be order $\delta^2$.
We then expand $d\sigma/dudt$  in powers of $\delta$.  
The leading terms from the tree diagram, the interference, and the loop diagrams  are
\begin{subequations}
\begin{eqnarray}
\frac{d\sigma_\mathrm{tree}}{du dt} &\approx& 
\frac{\alpha^3\nu^2}{256 M_{*0}^2} \, \frac{1}{t_-^2 + M_{*0}^2 \Gamma_{*0}^2} \,
\Bigg[   \big| A_0 \big|^2 \, D(s_-,u_-,t_-)
 \nonumber\\ 
&&\hspace{3cm}
 + \frac13 \left| A_0 -\frac{A_2}{\sqrt{5}} \right|^2 
\big[ 8 \delta^2 (s_- - 2t_-) - 3D(s_-,t_-,u_-) \big]
 \nonumber\\ 
&&\hspace{3cm}
 +  \frac{1}{20}\big| A_2 \big|^2
\big[ 24 \delta^2 (s_- \!-\! 2t_-) + 11 D(s_-,u_-,t_-) \big] 
\Bigg],
\label{sigmaDDgamma-tree}
\\
\frac{d\sigma_\mathrm{int}}{du dt} &\approx& 
 \frac{\pi  \alpha^3 \nu^2 \delta^2}{6M_{*0}^2}
\left( \left| A_0 - \frac{A_2}{\sqrt{5}} \right|^2  + \frac{9}{20} \big| A_2 \big|^2 \right)
 \big[- \delta^2 -(s_-  \!-\! u_- \!-\! 2t_-) \big] \nonumber\\ 
&&\hspace{1cm}
\times 
\mathrm{Re}\Bigg[ 
 \frac{M_X F(s,u)}{-\gamma_X - i \, \sqrt{u_- + 2 i M_{*0} \Gamma_{*0}}/2} 
 \left( \frac{1}{t_- + i M_{*0} \Gamma_{*0} }\right)^{\!\!*} \Bigg] ,
\label{sigmaDDgamma-int}
\\
\frac{d\sigma_\mathrm{loop}}{du dt} &\approx& 
\frac{8\pi^2 \alpha^3\nu^2 \delta^4}{3M_{*0}^2} \,
 \left( \left| A_0 - \frac{A_2}{\sqrt{5}} \right|^2  + \frac{9}{20} \big| A_2 \big|^2 \right)
\left|  \frac{M_X F(s,u)}{-\gamma_X - i \, \sqrt{u_- + 2 i M_{*0} \Gamma_{*0}}/2} \right|^2  .
\label{sigmaDDgamma-loop}
\end{eqnarray}
\label{sigmaDDgamma}%
\end{subequations}
Given that $M_XF(s,u)$ is order $1/\delta$, all three terms are order $\delta^0$.
For $t_-$ close to 0, the tree contribution is enhanced by the Breit-Wigner resonance factor.
The loop and interference contributions may be enhanced along the triangle singularity line $\sqrt{u_-} + \sqrt{s_-} = \delta$.

The integrals over  $t$ of the terms in Eqs.~\eqref{sigmaDDgamma} 
can be evaluated analytically to obtain the differential cross section $d\sigma/du$.
The Schmid cancellation can be exhibited by adding the interference and loop contributions:
\begin{eqnarray}
\frac{d\sigma_\mathrm{loop}}{du} + \frac{d\sigma_\mathrm{int}}{du} &\approx& 
 \frac{\pi  \alpha^3 \nu^2 \delta^4}{6M_{*0}^2}
 \left( \left| A_0 - \frac{A_2}{\sqrt{5}} \right|^2  + \frac{9}{20} \big| A_2 \big|^2 \right)
 \nonumber\\ 
&&\hspace{-3cm}
\times   \mathrm{Re}\Bigg[ 
 \frac{ M_X F(s,u)}{-\gamma_X - i \, \sqrt{u_- + 2 i M_{*0} \Gamma_{*0}}/2} 
\left( \frac{32 \pi  \delta\, M_XF(s,u)^* \,  \sqrt{u_-}}{-\gamma_X + i \, \sqrt{u_- - 2 i M_{*0} \Gamma_{*0}}/2} \right.
 \nonumber\\ 
&&\hspace{-2cm}
\left. + \frac{\delta^2 + s_--u_- - 2 i M_{*0} \Gamma_{*0}}{\delta^2}
 \log \frac{(\delta +\sqrt{u_-}\, )^2 -s_-  + 2 iM_{*0} \Gamma_{*0}}
  {(\delta -\sqrt{u_-}\,)^2 -s_-  + 2i M_{*0} \Gamma_{*0}}  + \frac{4\sqrt{u_-}}{\delta} \right)\Bigg].
\label{sigmaDDgamma/du-loop+int}
\end{eqnarray}
In the limit $\Gamma_{*0} \to 0$ and $\gamma_X \to 0$,
there is a cancellation between a logarithm in $F(s,u)^*$ whose argument has real part
$\sqrt{u_-} + \sqrt{s_-}- \delta$ and the corresponding explicit logarithm in the second factor inside the real part.
This is the  cancellation pointed out by Schmid \cite{Schmid:1967ojm}.
The  cancellation leaves a single logarithm of $\sqrt{u_-} + \sqrt{s_-}- \delta$ from the factor of $F(s,u)$,
in agreement with the analysis of Anisovich and Anisovich \cite{Anisovich:1995ab}.
That  logarithm gives a divergence in the limit $\Gamma_{*0} \to 0$.

An expression for the loop amplitude $F(s,u)$ that is accurate up to corrections of order $\delta/M_{*0}$
is given in Eq.~\eqref{FLorentz:intzetat}.
The contribution to the differential cross section from loops plus interference in Eq.~\eqref{sigmaDDgamma/du-loop+int}
can be simplified by taking the limits $\Gamma_{*0} \to 0$ and $\gamma_X \to 0$.
As $\sqrt{u_-}$ approaches the triangle singularity at $\delta - \sqrt{s_-}$, 
the diverging term is
\begin{eqnarray}
\frac{d\sigma_\mathrm{loop}}{du} + \frac{d\sigma_\mathrm{int}}{du} &\longrightarrow& 
 \frac{\alpha^3 \nu^2 \delta^2}{48M_{*0}^2}
 \left( \left| A_0 - \frac{A_2}{\sqrt{5}} \right|^2  + \frac{9}{20} \big| A_2 \big|^2 \right)
 \nonumber\\ 
&&\hspace{1cm}
\times   \left( - \frac{x^2\log\big((1-x)/x\big) +x}{1-x} \right)
\log \frac{2\delta}{\big| \sqrt{u_-} + \sqrt{s_-} - \delta \big|},
\label{sigmaDDgamma/du-loop+int:tri}
\end{eqnarray}
where $x = \sqrt{s_-}/\delta$.
There is a single logarithm that diverges at the triangle singularity. 
Its coefficient is negative for $x$ between 0 and  0.782 and positive for  $x$ between 0.782 and  1.
The diverging negative contribution to the  cross section is not a problem, 
because the tree contribution also has diverging terms.
It has terms that diverge as $1/\Gamma_{*0}$ as $\Gamma_{*0} \to 0$
as well as terms that diverge logarithmically as $u$ approaches the triangle singularity.  
The integral over $t$ of the tree contribution to the differential cross section  in Eq.~\eqref{sigmaDDgamma-tree}
can be simplified by taking the limit $\Gamma_{*0} \to 0$.
As $\sqrt{u_-}$ approaches the triangle singularity at $\delta - \sqrt{s_-}$, 
the diverging terms are
\begin{eqnarray}
\frac{d\sigma_\mathrm{tree}}{du} &\longrightarrow& 
 \frac{\alpha^3 \nu^2 \delta^2}{48M_{*0}^2}
\Bigg[  \left( \left| A_0 - \frac{A_2}{\sqrt{5}} \right|^2  + \frac{9}{20} \big| A_2 \big|^2 \right)
  \frac{\pi \, s_-}{2M_{*0}\Gamma_{*0}}\theta\big( \sqrt{u_-} + \sqrt{s_-} - \delta \big)
 \nonumber\\ 
&&\hspace{2cm}
+ \frac{3}{2}\left( (1-x) \big| A_0 \big|^2
-\frac{1 - 3x}{3} \left| A_0 - \frac{A_2}{\sqrt{5}} \right|^2  
+ \frac{17 - 11x}{20} \big| A_2 \big|^2 \right)
\nonumber\\ 
&&\hspace{6cm}
\times \log \frac{2\delta}{\big| \sqrt{u_-} + \sqrt{s_-} - \delta \big|} \Bigg].
\label{sigmaDDgamma/du-tree:tri}
\end{eqnarray}
The sum of the two contributions to $d\sigma/du$ in Eqs.~\eqref{sigmaDDgamma/du-loop+int:tri}
and \eqref{sigmaDDgamma/du-tree:tri} is positive definite.
Thus the coefficient of the logarithm in the full cross section is positive.

\begin{figure}[hbt]
\includegraphics*[width=0.7\linewidth]{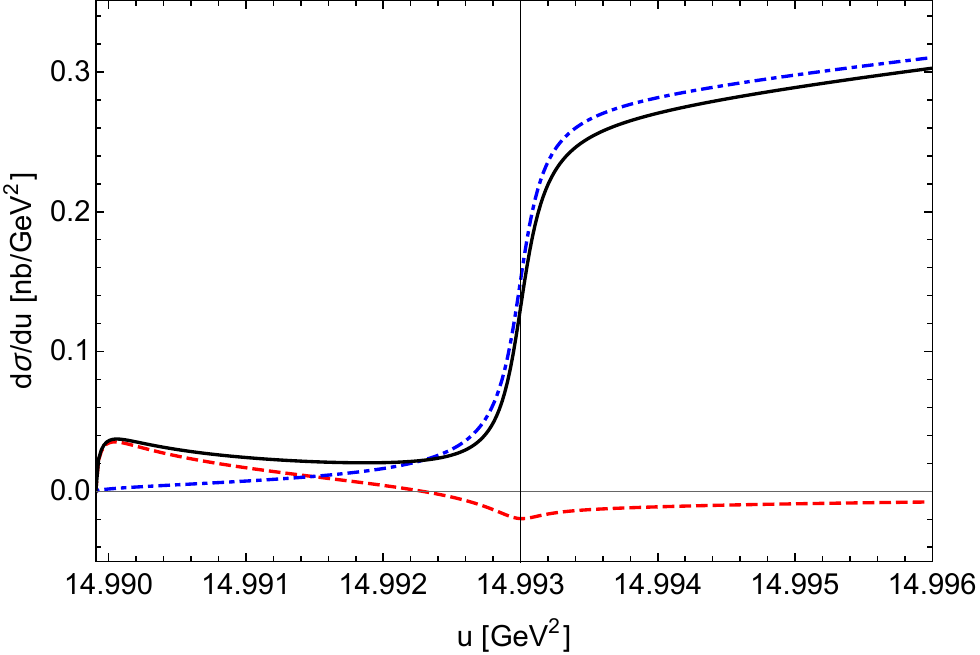} 
\caption{
Differential cross section $d\sigma/du$  for producing $D^{*0} \bar D^0 + \gamma$
integrated over $t$ as a function of  $u$ at $\sqrt{s}=4014.7$~MeV:
tree (dot-dashed blue curve), loop+interference (dashed red curve), and total (solid black curve).
The binding energy and width of $X$ are $|E_X| = 0.05$~MeV, $\Gamma_X = \Gamma_{*0}$.
The vertical  line is at the predicted value $u_\triangle=14.993~\mathrm{GeV}^2$ for the triangle singularity.
}
\label{fig:dsigma/du}
\end{figure}

In Fig.~\ref{fig:dsigma/du}, we illustrate the Schmid cancellation by showing 
the differential cross section $d\sigma/du$ integrated over $t$ as a function of $u$ 
for the physical $D^{*0}$ width $\Gamma_{*0} = 55$~keV.
We choose the center-of-mass energy  $\sqrt{s} = 4014.7$~MeV, which is 1~MeV
above the $D^{*0} \bar{D}^{*0} $ threshold and near the middle of the triangle-singularity region.
The triangle singularity is predicted by Eq.~\eqref{u-triangle} to be at $u_\triangle =14.993~\mathrm{GeV}^2$.
The tree contribution to $d\sigma/du$ and the contribution from loops plus interference are shown 
in Fig.~\ref{fig:dsigma/du} as well as their sum.
The tree contribution is strongly suppressed below $u_\triangle$.  
It is smaller than the contribution from loops plus interference for $u< 14.9915~\mathrm{GeV}^2$.
The tree contribution increases dramatically as $u$ increases past $u_\triangle$,
because the  $\bar{D}^{*0} $ resonance becomes kinematically accessible.
The contribution from loops plus interference is negative for $u > 14.9923~\mathrm{GeV}^2$,
and it has a small peak near the triangle singularity.
Since $\sqrt{s-4M_{*0}^2}/\delta =0.631$, which is less than 0.782, this negative peak is qualitatively consistent 
with our limiting expression in Eq.~\eqref{sigmaDDgamma/du-loop+int:tri}, 
which has a negative logarithmic divergence.
Our limiting expression for the tree contribution in Eq.~\eqref{sigmaDDgamma/du-tree:tri}  
includes a  term with a larger positive  logarithmic divergence.
For the physical $D^{*0}$ width, a term from the tree contribution with a positive peak 
whose height is comparable to that of the negative peak  from loops plus interference in Fig.~\ref{fig:dsigma/du}
would not be visible.
It would be completely overwhelmed by the rapidly increasing contribution 
from  the opening up of the $\bar{D}^{*0} $ resonance.

\begin{figure}[hbt]
\includegraphics*[width=0.7\linewidth]{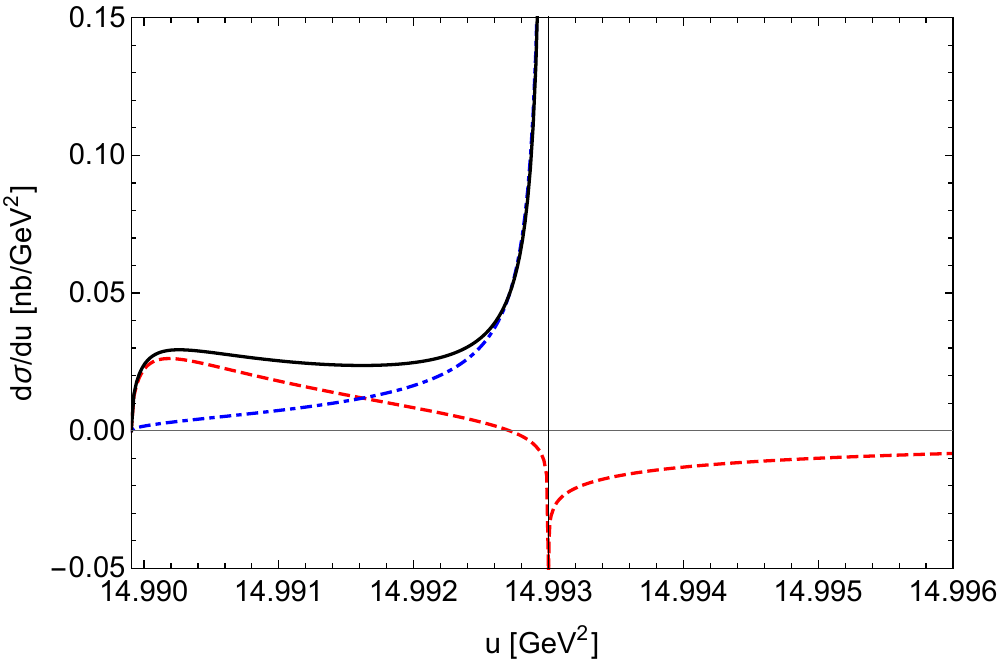} 
\caption{
Differential cross section $d\sigma/du$  for producing $D^{*0} \bar D^0 + \gamma$
integrated over $t$ as a function of  $u$ at $\sqrt{s}=4014.7$~MeV:
tree (dot-dashed blue curve), loop+interference (dashed red curve), and total (solid black curve).
The binding energy and width of $X$ are $|E_X| = 0.05$~MeV, $\Gamma_X =0$.
The vertical  line is at the predicted value $u_\triangle=14.993~\mathrm{GeV}^2$ for the triangle singularity.
The tree contribution to  $d\sigma/du$ is infinitely large for $u$ beyond the triangle singularity.
 }
\label{fig:dsigma/du:Gamma*0=0}
\end{figure}

In Fig.~\ref{fig:dsigma/du:Gamma*0=0}, we show the differential cross section $d\sigma/du$ integrated over $t$ 
as a function of $u$ in the limit $\Gamma_{*0} \to 0$ for $\sqrt{s} = 4014.7$~MeV.
The tree contribution is  infinitely large for $u > u_\triangle$,
because it includes an integral over $t$ of a Breit-Wigner resonance with zero width.
The contribution from loops plus interference is negative near the triangle singularity,
in agreement with the limiting expression in Eq.~\eqref{sigmaDDgamma/du-loop+int:tri}.
It diverges at $u=u_\triangle$, with a shape consistent with a negative single log.
The limiting expression for the tree contribution to $d\sigma/du$
in Eq.~\eqref{sigmaDDgamma/du-tree:tri} suggests that it  also has a term that
diverges at $u=u_\triangle$, with a shape consistent with a positive single log.
Its effects are not visible for $u < u_\triangle$, because  other terms in the tree contribution are growing even more rapidly.

\begin{figure}[hbt]
\includegraphics*[width=0.7\linewidth]{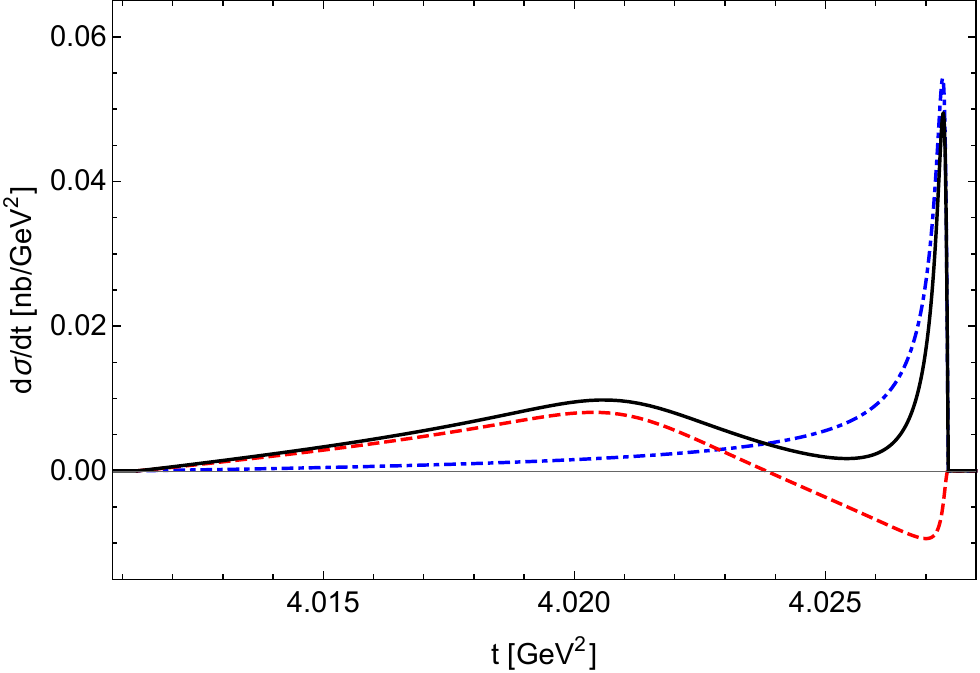} 
\caption{
Differential cross section $d\sigma/dt$  for producing $D^{*0} \bar D^0 + \gamma$
integrated over $u<u_\triangle(s)$ as a function of  $t$ at $\sqrt{s}=4014.7$~MeV:
tree (dot-dashed blue curve), loop+interference (dashed red curve), and total (solid black curve).
The binding energy and width of $X$ are $|E_X| = 0.05$~MeV, $\Gamma_X = \Gamma_{*0}$.
 }
\label{fig:dsigma/dt:min}
\end{figure}

For the physical value of the $D^{*0}$ width, the triangle singularity is not directly observable
as a peak in $d\sigma/dudt$ at a value of $t$ below the $\bar D^{*0}$ resonance band
or as a peak in $d\sigma/du$ integrated over $t$. 
It is possible however that  the triangle singularity could be observed indirectly in some other way.
In the Dalitz plot in Fig.~\ref{fig:Dalitzcorner},  the density of points 
is significantly larger to the left of the triangle-singularity line than just to the right of that line.
However the density of points to the left of the triangle-singularity line decreases 
as $t$ approaches the $\bar D^{*0}$ resonance band.
This decrease in the density of points  for $t$ near the resonance band
arises from an interference effect associated with the triangle singularity.
This effect can be quantified by calculating $d\sigma/dt$ integrated over $u<u_\triangle(s)$ as a function of $t$.
This function is shown in Fig.~\ref{fig:dsigma/dt:min} for $\sqrt{s} = 4014.7$~MeV.
The tree contribution behaves as $|t_-|/(t_-^2 + M_{*0}^2  \Gamma_{*0}^2)$ near the upper kinematic endpoint
at $t = M_{*0}^2$, which is the center of the $\bar D^{*0}$ resonance band.
The factor of $|t_-|$ comes from the integration over $u$.  The tree contribution
increases to a sharp peak near $t_- = - M_{*0}  \Gamma_{*0}$, before decreasing to 0 at $t_-=0$.
The contribution from loops plus interference is negative for $t$ larger than about 4.024~GeV$^2$.
It decreases to a minimum near 4.027~GeV$^2$, before increasing sharply to 0.
The cancellation between the tree contribution and the contribution from loops plus interference
produces a local minimum near 4.025~GeV$^2$.

\begin{figure}[hbt]
\includegraphics*[width=0.7\linewidth]{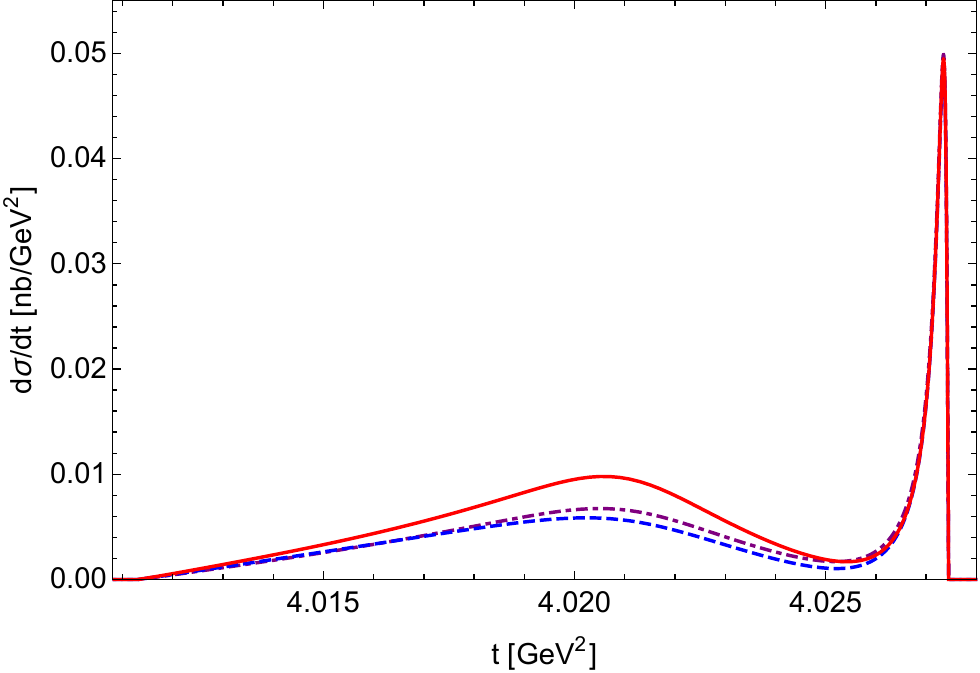} 
\caption{
Differential cross section $d\sigma/dt$ for producing $D^{*0} \bar D^0 + \gamma$
integrated over $u<u_\triangle(s)$ as a function of  $t$ at $\sqrt{s}=4014.7$~MeV.
The binding energy and width of $X$ are 
$|E_X| = 0.05$~MeV, $\Gamma_X = \Gamma_{*0}$ (solid red curve),
$|E_X| = 0.10$~MeV, $\Gamma_X = \Gamma_{*0}$ (dashed blue curve), 
and $|E_X| = 0.05$~MeV, $\Gamma_X = 2\, \Gamma_{*0}$  (dot-dashed purple curve).
 }
\label{fig:dsigma/dt:min2}
\end{figure}

 The local minimum in the differential cross section $d\sigma/dt$ integrated over
$u<u_\triangle(s)$ is insensitive to the binding energy and the width of $X$.  
This is illustrated in Fig.~\ref{fig:dsigma/dt:min2}, which shows
the cross section for the first three choices of $E_X$, $\Gamma_X$
enumerated after Eq.~\eqref{EX,GammaX}.  For $|E_X| = 0.05$~MeV,  $\Gamma_X = \Gamma_{*0}$,
the  shallow minimum in $d\sigma/dt$ is at $t =4.0254~\mathrm{GeV}^2$.
The contrast defined by the ratio of the value at the local maximum at $t =4.0206~\mathrm{GeV}^2$ and the value at the local minimum is 5.8.
The narrow peak near the upper endpoint of $t$ is almost unaffected by changing 
either $|E_X|$ or $\Gamma_X$ by a factor of 2.
Doubling either $|E_X|$ or $\Gamma_X$ decreases the position of the local minimum in $t$ by $0.0002~\mathrm{GeV}^2$.
Doubling $|E_X|$ decreases the contrast between the local maximum and the local minimum to 5.6,
while doubling $\Gamma_X$ decreases  the contrast  to 3.9.


\section{Summary}
\label{sec:Summary}

If $X(3872)$ is a weakly bound charm-meson molecule,
a charm-meson triangle singularity produces a narrow peak in the cross section for 
$e^+ e^-$ annihilation into $X+\gamma$ just above the $D^{*0} \bar D^{*0}$ threshold \cite{Braaten:2019gfj}.
The peak appears near  a value of the center-of-mass energy $\sqrt{s}$ given by $s_\triangle$ in Eq.~\eqref{s-triangle}.
The cross section is shown as a function of $\sqrt{s}$ in Fig.~\ref{fig:sigmaXgamma}
for several values of the binding energy $|E_X|$ and the width $\Gamma_X$ of $X$.
The observation of the narrow peak in the cross section would support the identification of $X$
as a weakly bound charm-meson molecule.

In this paper, we have studied the effects of the charm-meson triangle singularity on the cross section for 
$e^+ e^-$ annihilation into $D^{*0} \bar D^0 +\gamma$ just above the $D^{*0} \bar D^{*0}$ threshold.
The matrix element is the sum of the matrix element $\mathcal{M}_\mathrm{tree}$ in Eq.~\eqref{MDDgamma-tree}
from the tree diagram and  the matrix element $ \mathcal{M}_\mathrm{loop}$ in Eq.~\eqref{MDDgamma-loop}
 from the  loop diagrams.
The differential cross section $d\sigma/dudt$ is a function of the  squared invariant masses 
$u$ of $D^{*0} \bar D^0$ and $t$ of $\bar D^0 \gamma$ as well as the squared center-of-mass energy $s$.
The matrix element $ \mathcal{M}_\mathrm{loop}$ has a factor of the loop amplitude $F(s,u)$,
which depends on the $D^{*0}$  width $\Gamma_{*0}$,
and it also has a denominator that depends on the complex binding momentum $\gamma_X$ of $X$. 
The loop amplitude $F(s,u)$ is expressed as a Lorentz-invariant integral in Eq.~\eqref{FLorentz} of Appendix~\ref{LoopIntegralR}.
A nonrelativistic approximation to the loop amplitude is given analytically in Eq.~\eqref{FanalyticNR} of Appendix~\ref{LoopIntegralNR}.
If the width $\Gamma_{*0}$ of the $D^{*0}$ is sufficiently small,
the triangle singularity produces a narrow peak in the differential cross section $d\sigma/dudt$ 
as a function of $u$ near the value $u_\triangle(s)$ in Eq.~\eqref{u-triangle}, 
which requires $t$ to be below the $\bar D^{*0}$ resonance band,  for $s$ in the narrow range between 
the $D^{*0} \bar D^{*0}$ threshold and $s_\triangle$ in Eq.~\eqref{s-triangle}.

The effects of the triangle singularity are particularly dramatic in the limit $\Gamma_{*0} \to 0$.
The differential cross section $d\sigma/dudt$ has a term proportional to $\log^2|u-u_\triangle(s)|$ 
that diverges at $u=u_\triangle(s)$.
As pointed out by Schmid, there is a cancellation of the $\log^2$ divergence 
in $d\sigma/du$ integrated over $t$ \cite{Schmid:1967ojm}.  
The cancellation is between $ |\mathcal{M}_\mathrm{loop}|^2$ along the triangle singularity line
and the interference term
$2 \,\mathrm{Re}[\mathcal{M}_\mathrm{loop} \mathcal{M}_\mathrm{tree}^*]$ 
in the region near the resonance band.
The exact cancellation of the $\log^2$ terms requires both $\Gamma_{*0} \to 0$ and $\gamma_X \to 0$.
The Schmid cancellation leaves a term in $d\sigma/du$ 
 proportional to $\log|u-u_\triangle(s)|$, as pointed out by Anisovich and Anisovich \cite{Anisovich:1995ab}.
The position of that divergence coincides with the dramatic increase in $d\sigma/du$
proportional to $1/\Gamma_{*0}$ from the $\bar D^{*0}$ resonance becoming kinematically accessible.

For the physical value of the $D^{*0}$ width, the triangle singularity is not directly observable
as a peak in a differential cross section.  It is not observable as a peak in $d\sigma/dudt$, because
the suppression from  the $D^{*0}$ width reduces the peak in $u$ near $u_\triangle(s)$
to a shoulder that drops off beyond $u_\triangle(s)$, as illustrated in Fig.~\ref{fig:dsigma/dudt}.
The triangle singularity is not observable as a peak in $d\sigma/du$ integrated over $t$, 
because the term with a peak near $u=u_\triangle(s)$ 
is overwhelmed by the rapidly increasing tree contribution, as illustrated in Fig.~\ref{fig:dsigma/du}.

The effects of the charm-meson triangle singularity may be observable indirectly as a local minimum
in $d\sigma/dt$ integrated over $u<u_\triangle(s)$.
The differential cross section is shown as a function of $t$ in Fig.~\ref{fig:dsigma/dt:min2}
for several values of $|E_X|$, $\Gamma_X$.
The local minimum comes from destructive interference between 
$|\mathcal{M}_\mathrm{loop}|^2+2 \,\mathrm{Re}[\mathcal{M}_\mathrm{loop} \mathcal{M}_\mathrm{tree}^*]$
and $|\mathcal{M}_\mathrm{tree}|^2$ that is related to the Schmid cancellation.
The observation of this local minimum would provide additional support for the identification of $X$
as a weakly bound charm-meson molecule.

\begin{acknowledgments}
This work was supported in part by the U.S.\ Department of Energy under grant DE-SC0011726,
the National Natural Science Foundation of China under grant 11905112, 
the Natural Science Foundation of Shandong Province under grant ZR2019QA012, 
and the Fundamental Research Funds of Shandong University under grant 2019GN038. 
Jun Jiang thanks the China Scholarship Council for their support.
\end{acknowledgments}


\appendix

\section{Nonrelativistic  Loop Integral} 
\label{LoopIntegralNR}

In Ref.~\cite{Braaten:2019gwc}, the loop diagrams for $e^+ e^- \to X+ \gamma$ in Fig.~\ref{fig:eetogammaX}
were evaluated in the center-of-momentum (CM) frame using nonrelativistic charm-meson propagators
and nonrelativistic vertices. 
The loop amplitude can be obtained from a more general loop amplitude $F(W,U)$ for the production 
of  $D^{*0} \bar{D}^0 + \gamma$ through the loop diagrams in Fig.~\ref{fig:eetogammaDDloop}.
That amplitude is a function of  the   total energy  $W=\sqrt{s} - 2M_{*0}$ 
relative to the $D^{*0}\bar{D}^{*0}$ threshold in the $e^+ e^-$ CM frame
and the energy $U = \sqrt{u} -(M_{*0} + M_0)$
of the $D^{*0} \bar{D}^0$ pair relative to their threshold in their CM frame.
These two independent energies are related by energy conservation 
to the photon energy $q$ in the $e^+ e^-$ CM frame:
\begin{equation}
W =  (q-\delta)  + U + q^2/(2M_X),
 \label{energyconserve} 
\end{equation}
where $\delta = M_{*0} - M_0$.
The loop amplitude can be
expressed in terms of a nonrelativistic loop integral that is ultraviolet convergent:\footnote
{The normalization of $F(W,U)$  differs from that of $F(W)$  in Ref.~\cite{Braaten:2019gwc}
by a constant  factor $1/(8M_{*0}\sqrt{\pi \gamma_X}\,)$ and by a factor $1/q^2$ that depends on the photon energy.}
\begin{eqnarray}
 F(W,U)&=& 
-  \frac{i}{8M_{*0}^2 M_0}
    \int\!\! \frac{d^3k}{(2\pi)^3}\, \frac{\bm q \cdot \bm k}{\bm q^2} \! \int\! \frac{d\omega}{2\pi} 
\, \frac{1}{\omega - \bm k^2/(2M_{*0})+ i \Gamma_{*0}/2}
\nonumber\\
&& \hspace{-1cm} \times
   \frac{1}{W -\omega - \bm k^2/(2M_{*0})+ i\Gamma_{*0}/2}\,
   \frac{1}{W - (|\bm q| - \delta)  - \omega - (\bm q + \bm k)^2/(2M_0) + i \epsilon},
\label{Fdwd3k} 
\end{eqnarray}
where $\bm q$ is the 3-momentum of the photon  in the $e^+e^-$ CM frame.
The scalar integral in Eq.~\eqref{Fdwd3k} was obtained from a vector  integral  with a factor of $\bm{k}$ in the numerator
by an angle average that replaces that  vector  by $(\bm k \cdot \bm q/\bm q^2)\, \bm{q}$.
The loop  integral  in Eq.~\eqref{Fdwd3k} was evaluated analytically in Ref.~\cite{Braaten:2019gwc}:
\begin{equation}
F(W,U) =  \frac{i}{64\pi M_X q}
 \left(- \frac{b}{c}    \log\frac{\sqrt{a} +\sqrt{a+b+c} +  \sqrt{c}}{\sqrt{a} +\sqrt{a+b+c}  - \sqrt{c}}
-2\frac{\sqrt{a}  -\sqrt{a+b+c}}{\sqrt{c}}  \right) ,
\label{FanalyticNR}
\end{equation}
where $M_X =M_{*0} \!+\! M_0$, $q$ is the function of the difference $W\!-\!U$ 
determined by Eq.~\eqref{energyconserve},
and the coefficients $a$, $b$, and $c$ are
\begin{subequations}
\begin{eqnarray}
a &=&   M_{*0} W+i M_{*0}\Gamma_{*0} ,
    \\
b &=&  -\big[  (\mu/M_0)^2   q^2 +   M_{*0} W - 2 \mu U \big] - i  (\mu/M_0) M_{*0}\Gamma_{*0} ,
\label{coeffbNR}
    \\
c &=&  (\mu/M_0)^2  q^2.
\end{eqnarray}
\label{coefficientsabcNR}%
\end{subequations}
Note that the sum of the three coefficients  does not depend on $W$:
\begin{equation}
a+b+c = 2 \mu U + i \mu \Gamma_{*0}.   
 \label{a+b+cNR}
\end{equation}

In the case of production of $X+ \gamma$, the energy conservation condition is obtained by replacing $U$ 
in Eq.~\eqref{energyconserve} by the negative energy $E_X$ of $X$.
In Ref.~\cite{Braaten:2019gwc}, the loop amplitude in Eq.~\eqref{FanalyticNR}
was obtained by setting $U = E_X$ in Eq.~\eqref{a+b+cNR}.  
A more accurate result can be
obtained by analytically continuing $U$ to the complex energy $E_X-i\Gamma_X/2$ of $X$.
This changes the imaginary part of $a+b+c$ to  $-i\mu ( \Gamma_X -  \Gamma_{*0})$.
The change in sign of the imaginary part requires careful treatment of the branch cuts 
to ensure the continuity of $F(W, E_X \!-\! i\Gamma_X/2)$ as a function of $W$.
The factor $\Gamma_X -  \Gamma_{*0}$ in the imaginary part 
of $a+b+c$ can be interpreted as the partial width of the $X$ bound state 
into decay channels other than $D^0\bar D^0 \pi^0$ and $D^0\bar D^0 \gamma$.

The triangle singularity arises from the vanishing of the denominator of the argument of the logarithm in 
Eq.~\eqref{FanalyticNR}.  In the limit $\Gamma_{*0} \to 0$, this condition reduces to
\begin{equation}
\sqrt{M_{*0} W} + \sqrt{2 \mu U} =  (\mu/M_0)\, q.
 \label{triangleNR}
\end{equation}
We can get an equation relating $U$ and $W$ at the triangle singularity 
by solving this equation for $q$ and inserting it into the energy conservation condition in Eq.~\eqref{energyconserve}.
If we set $M_0=M_{*0}-\delta$ and take $M_{*0} W$ to be order $\delta^2$, 
this equation can be solved for $2 \mu U$ as an expansion in 
powers of  $\delta/M_{*0}$ beginning at order $\delta^2$: 
\begin{equation}
2\mu\, U_\triangle(W) = \frac14 \big(2 \sqrt{M_{*0} W}\, - \delta \big)^2 
+ \frac{\delta}{M_X} \big( \sqrt{M_{*0} W}\,\delta  - 2 M_{*0} W \big) +\ldots.
\label{Utri-W}
\end{equation}
This expansion agrees with that of the relativistic result for $u_\triangle(s)$ in Eq.~\eqref{u-triangle}
through order $\delta^3$,  but it disagrees at order $\delta^4$.


\section{Relativistic Loop Integral} 
\label{LoopIntegralR}

The loop diagrams for $e^+ e^- \to D^{*0} \bar{D}^0+ \gamma$ are shown in Fig.~\ref{fig:eetogammaDDloop}.
If these diagrams are evaluated using relativistic charm-meson propagators and relativistic vertices,
the loop integral reduces to
\begin{equation}
 \int \!\! \frac{d^4k}{(2\pi)^4} \frac{k_\nu (Q-k)_\alpha (g_{\lambda\rho} - k_\lambda k_\rho/M_{*0}^2)}
 {(k^2-M_{*0}^2 + i \epsilon) \, [(Q-k)^2-M_{*0}^2 + i \epsilon]\,  [(P-k)^2-M_0^2 + i \epsilon]},
\label{tensorintd4k} 
\end{equation}
where $Q=P+q$ is the total momentum, $P$ is the momentum of $D^{*0} \bar{D}^0$, 
$q$ is the momentum of $\gamma$, and $k$ is the loop momentum of the 
virtual spin-1 charm meson line attached to the elastic scattering vertex.
The momentum-dependent term in the numerator of the propagator for the other virtual spin-1 charm meson 
is eliminated by the contraction with the radiative transition vertex.
The decay width of the $D^{*0} $ has been set to zero 
to make the $D^{*0}$ propagators as simple as possible.  The decay width
can be reintroduced by the substitution $M_{*0}^2 -i \epsilon \to M_{*0}^2 - i M_{*0} \Gamma_{*0}$.
The loop integral in Eq.~\eqref{tensorintd4k}  is a 4-index tensor function of $Q$ and $q$.
It is contracted with the tensor $\mathcal{A}(Q)^{\mu \nu \lambda \sigma}$ in Eq.~\eqref{A[eetoD*D*]}
from the $\gamma^*$-to-$D^{*0} \bar D^{*0}$ vertex,
which sets terms with a factor of $Q_\nu$ or $Q_\lambda$ to 0.
It is also contracted with a tensor $\epsilon_\sigma^{~\alpha \beta \tau} q_\beta$ from the $D^{*0}$-to-$D^0 \gamma$ vertex,
which sets terms with a factor of $q_\alpha$ to 0.

The loop integral in Eq.~\eqref{tensorintd4k}  produces a triangle singularity determined by the following conditions. 
First, the three charm mesons that form the loop are all  on their mass shells: 
$k^2 = M_{*0}^2$, and $(Q-k)^2 = M_{*0}^2$, and
\begin{equation}
0=(P-k)^2-M_0^2 = u -2 \big[P_0\, k_0- |\bm{P}|\,  |\bm{k}|\, \cos\theta \big]+M_{*0}^2 - M_0^2,
\label{trianglePk}
\end{equation}
where $u=P^2$ and $\theta$ is the angle between the 3-momenta $\bm{P}$ and $\bm{k}$.
In the $e^+ e^-$ CM frame, we have $P_0=(s+u)/(2\sqrt{s})$,  $k_0=\sqrt{s}/2$,
$ |\bm{P}| = (s-u)/(2\sqrt{s})$,  and $|\bm{k}|=\sqrt{s-4M_{*0}^2}/2$, where $s=Q^2$.
A second condition is that the 3-momenta $\bm{P}-\bm{k}$ and $\bm{k}$ of the two charm mesons 
are parallel in the $e^+ e^-$ CM frame: $\cos\theta =1$. 
With this value of $\cos\theta$,  Eq.~\eqref{trianglePk} can be reduced to
\begin{equation}
-\frac{s-u}{2} \left[ 1- \sqrt{(s-4M_{*0}^2)/s}\,  \right] +(M_{*0}^2 - M_0^2)=0.
\label{triangle-su:equation}
\end{equation}
Solving for $u$ gives the expression for its value $u_\triangle(s)$ at the triangle singularity in Eq.~\eqref{u-triangle}.
This is a necessary condition for the triangle singularity, but it is not sufficient.
The final condition for the  triangle singularity is that the velocity of the spin-0 charm meson 
must be greater than or equal to that of
the spin-1 meson in order for it to overtake that meson so they can scatter: 
\begin{equation}
    \frac{|\bm P - \bm k|}{P_0 - k_0} \geq \frac{|\bm k|}{k_0}.
\label{triangle-velocity}
\end{equation}
This gives an upper bound on $u$ in terms of $s$:
\begin{equation}
    u \leq  \frac{s \big(\sqrt{s}-\sqrt{s-4M_{*0}^2}\,\big)^2}{4M_{*0}^2} .
\label{triangle-u}
\end{equation}
The minimum value of $u$ is $(M_{*0}+M_0)^2$.
Inserting this value into Eq.~\eqref{triangle-u} and solving for $s$ gives the upper endpoint
$s \le s_\triangle$  of the triangle-singularity region, where $s_\triangle$ is given in Eq.~\eqref{s-triangle}.

The differential cross sections for  $e^+ e^- \to X+ \gamma$ in Eq.~\eqref{dsigma/dOmega:Xgamma}
and $e^+ e^- \to D^{*0} \bar{D}^0+ \gamma$ in Eq.~\eqref{sigmaDDgamma}
are expressed in terms of a Lorentz-invariant loop amplitude $F(s,u)$.
The nonrelativistic approximation to this amplitude is given as a loop integral in Eq.~\eqref{Fdwd3k}.
A  Lorentz-invariant generalization of that amplitude
can be obtained by replacing the nonrelativistic propagators in Eq.~\eqref{Fdwd3k} 
by the appropriate relativistic propagators 
and replacing the multiplicative  factor of  $\bm{q} \cdot \bm{k}$
by the appropriate Lorentz-invariant generalization:
\begin{equation}
F(s,u) = -\frac{i}{(Q.q)^2} \int \!\! \frac{d^4k}{(2\pi)^4} \frac{Q.q\, Q.k - Q^2\, q.k}
 {(k^2-M_{*0}^2 + i \epsilon) \, [(Q-k)^2-M_{*0}^2 + i \epsilon]\,  [(P-k)^2-M_0^2 + i \epsilon]},
\label{FLorentz}
\end{equation}
where $Q^2=s$, $P^2=u$, and $Q.q=(s-u)/2$.
This can be expressed as an integral over two Feynman parameters:
\begin{equation}
F(s,u) = - \frac{1}{16\pi^2} 
 \int_0^1\!\!dx   \int_0^{1-x}\!\!\!\!dy\,\frac{y}{(1 \!-\! x \!-\! y)(xs + yu) -(1\!-\!y) M_{*0}^2 - yM_0^2 + i \epsilon }.
\label{FLorentz:Fp}
\end{equation}
The nonrelativistic loop amplitude $F(W,U)$ in Eq.~\eqref{Fdwd3k}, with $W= \sqrt{s}-2M_{*0}$
and $U= \sqrt{s}-(M_{*0}\!+\!M_0)$, gives an excellent approximation to the function $F(s,u)$ 
in the triangle singularity region.
In the region where $\sqrt{s}$ is less than 10~MeV above $2M_{*0}$
and $\sqrt{u}$ is less than 10~MeV above $M_{*0}\!+\!M_0$, 
the numerical differences seem to be consistent with errors of order $(\delta/M_{*0})^2$. 

We proceed to verify analytically that the relativistic loop amplitude $F(s,u)$ in Eq.~\eqref{FLorentz}
can be reduced to the nonrelativistic loop amplitude $F(W,U)$ in Eq.~\eqref{Fdwd3k}.
We will verify this only up to errors first order in $\delta/M_{*0}$.
It is convenient to change the integration variables in Eq.~\eqref{FLorentz:Fp} to $t$ and $\zeta$ defined by
$x= \tfrac12(1+\zeta)(1-t)$ and $y=\tfrac12(1+\zeta)t$.  
In the region where $s-4M_{*0}^2$ and $u-M_X^2$ are both order $\delta^2$,
the leading term in the expansion of $F(s,u)$  in powers of $\delta/M_{*0}$
comes from the region where $\zeta M_X$ is order $\delta$.
If we keep only the leading terms in the numerator and in the denominator, the integral reduces to
\begin{eqnarray}
F(s,u) &\approx& \frac{1}{32\pi^2} 
 \int_0^1 dt \int_{-1}^{+1} d\zeta  
\nonumber\\
&& \hspace{-1cm} \times 
 \frac{t}
 {(\zeta M_X-t\delta)^2+  t (1 \!-\! t) \delta^2 - (1 \!-\! t)(s-4M_{*0}^2) - t (u-M_X^2)  - 2 i (2-t) M_{*0}\Gamma_{*0} },
\nonumber\\
\label{FLorentz:FLO}
\end{eqnarray}
where $M_X = M_{*0} \!+\! M_0$.
We have reintroduced the width of the $D^{*0}$  through the substitution 
$M_{*0}^2 -i \epsilon \to M_{*0}^2 - i M_{*0} \Gamma_{*0}$ 
and treated $M_{*0}\Gamma_{*0}$ as order  $\delta^2$.
Since the leading contributions to the integral in Eq.~\eqref{FLorentz:FLO}
come from the region where $M_X\zeta$ is order $\delta$,
the endpoints of the integral over $\zeta$ can be extended to $\pm \infty$.
The analytic result for the integral over $\zeta$ then becomes relatively simple:
\begin{equation}
F(s,u) \approx  \frac{1}{32\pi M_X}  \int_0^1 dt 
\frac{t}{\sqrt{t(1 \!-\! t) \delta^2 - (1 \!-\! t) (s - s_0) - t\, (u - u_0) }} ,
\label{FLorentz:intzeta}
\end{equation}
where $s_0 = 4M_{*0}^2 - 4i  M_{*0}\Gamma_{*0} $ and $u_0 =  (M_{*0}+M_0)^2 - 2 i M_{*0}\Gamma_{*0}$. 
The final integral over $t$ can also be evaluated analytically:
\begin{eqnarray}
F(s,u)& \approx& \frac{i}{64\pi M_X\, \delta} 
\Bigg( \frac{\delta^2 + (s - s_0) - (u - u_0)}{\delta^2}
\log \frac{\sqrt{s - s_0} + \sqrt{u - u_0} +\delta}{\sqrt{s - s_0} + \sqrt{u - u_0}-\delta} 
\nonumber\\
&& \hspace{5cm}
  - 2 \frac{\sqrt{s - s_0} - \sqrt{u - u_0}}{\delta} \Bigg).
\label{FLorentz:intzetat}
\end{eqnarray}
If we  make the substitutions $s \to  (2M_{*0}+W)^2$ and $u \to (M_{*0}\!+\!M_0 +U)^2$,
 this agrees to leading order in $\delta/M_{*0}$ 
with the nonrelativistic loop amplitude $F(W,U)$ in Eq.~\eqref{Fdwd3k} 
in the region where $s-4M_{*0}^2$ and $u-(M_{*0}\!+\!M_0)^2$ are order $\delta^2$.
The analytic approximation to $F(s,u)$  in Eq.~\eqref{FLorentz:intzetat}  has errors that are first order in $\delta/M_{*0}$,
so it is not useful as a quantitative approximation.
The nonrelativistic amplitude $F(W,U)$  in Eq.~\eqref{FanalyticNR} is a much better approximation. 

We proceed to show that the tensor loop integral in Eq.~\eqref{tensorintd4k}
 can be reduced to the scalar loop integral in Eq.~\eqref{FLorentz}
in the region where $s-4M_{*0}^2$ and $u-M_X^2$ are both order $\delta^2$.
We will verify this only to leading order in $\delta/M_{*0}$.
In a frame where $q^\mu$ is order $\delta$,
the tensor loop integral in Eq.~\eqref{tensorintd4k} can be expanded in powers of $\delta/M_{*0}$
with $M_0=M_{*0} - \delta$.
The loop integral  in Eq.~\eqref{tensorintd4k} is quadratically ultraviolet divergent.
Up to corrections suppressed by $\delta/M_{*0}$,
the tensor $g_{\lambda\rho} - k_\lambda k_\rho/M_{*0}^2$ in the numerator of Eq.~\eqref{tensorintd4k} 
can be  replaced by $g_{\lambda\rho}-Q_\lambda Q_\rho/Q^2$
and then pulled outside the loop integral.
This reduces the numerator of the loop integral in Eq.~\eqref{tensorintd4k}  to $k_\nu (Q-k)_\alpha$,
so the ultraviolet divergence  is only logarithmic.
That loop integral is a tensor function of the 4-momenta $Q$ and $q$ with indices $\nu$ and $\alpha$.
It can be decomposed into a linear combination of the 5 tensors
$g_{\nu \alpha}$, $Q_\nu Q_\alpha$, $Q_\nu q_\alpha$, $q_\nu Q_\alpha$, and $q_\nu q_\alpha$
with coefficients that are functions of $Q^2$ and $Q.q$.
Tensors with a factor of $Q_\nu$ contract to 0 with the $\gamma^*$-to-$D^{*0} \bar D^{*0}$ vertex.
Tensors with a factor of $q_\alpha$ contract to 0 with the $\bar D^{*0}$-to-$\bar D^0\gamma$ vertex.
The only tensors that remain are $g_{\nu \alpha}$ and $q_\nu Q_\alpha$.
The coefficient of $g_{\nu \alpha}$ is suppressed compared to that of $q_\nu Q_\alpha$ by a factor of 
order $\delta^2$. Since $q_\nu$ is order $\delta$, the $g_{\nu \alpha}$ term is suppressed compared to 
the $q_\nu Q_\alpha$ term by a factor of order $\delta$.

The reduction of the tensor loop integral in Eq.~\eqref{tensorintd4k} to the scalar loop integral in Eq.~\eqref{FLorentz}
 can be completed by showing that the numerator factor $k_\nu (Q-k)_\alpha$
can be replaced by the numerator factor in Eq.~\eqref{FLorentz}
multiplied by $\tfrac12  q_\nu Q_\alpha/(Q.q)^2$ up to corrections suppressed by factors of $\delta/M_{*0}$.
In the tensor reduction of the momentum integral of the $k_\nu (Q-k)_\alpha$ term, 
the scalar integral coefficient of $q_\nu Q_\alpha$  term is very different
from half the scalar momentum integral in Eq.~\eqref{FLorentz}.  
For example, its numerator  includes terms quadratic in the loop momentum $k$.
However after introducing Feynman parameters and integrating over the loop momentum $k$,
 the Feynman parameter integral differs from that in Eq.~\eqref{FLorentz:Fp} only by a factor of $1-x-y$ in the numerator.
After the change of variables to $t$ and $\zeta$ as in Eq.~\eqref{FLorentz:FLO},
that factor becomes $\tfrac12 (1-\zeta)$. 
Since the leading contribution comes from the region where 
$\zeta M_X$ is order $\delta$, that factor reduces to $\tfrac12$.
This completes the demonstration that the tensor loop integral 
can be reduced to the loop amplitude $F(s,u)$ in Eq.~\eqref{FLorentz}
up to corrections suppressed by at least one power of $\delta/M_{*0}$.

In the sum of the two triangle diagrams, the triangle of charm-meson propagators and the three attached vertices
can be represented by a Lorentz-covariant effective vertex.
The effective vertex for the coupling of  a virtual photon with momentum $Q=P\!+\!q$ to $X$ with momentum $P$
 and a real photon with momentum $q$ is
\begin{equation}
- 4ie^2 \nu  (2M_X)^{3/2}  \sqrt{\pi \gamma_X} \, F(Q^2,P^2)\,
\mathcal{A}(Q)^{\mu \nu \lambda \sigma} q_\nu \, 
\epsilon_{\sigma\alpha \beta \tau} Q^\alpha q^\beta ,
\label{effvertex:Xgamma}
\end{equation}
where $\mu$, $\lambda$, and $\tau$ are the Lorentz indices of the virtual photon, $X$, and real photon, respectively.
The effective vertex for the coupling of  a virtual photon with momentum $Q=P+q$
to $D^{*0}$ and $\bar D^0$ with total momentum $P$ and a real photon with momentum $q$ is
\begin{eqnarray}
&&- 8 i \pi  e^2 \nu M_X \frac{ \sqrt{4M_{*0}M_0}}
{-\gamma_X - i \, \lambda^{1/2}(P^2,M_{*0}^2 \!-\! i M_{*0} \Gamma_{*0}, M_0^2)/\big(2 \sqrt{P^2}\, \big)}
\nonumber\\
&& \hspace{2cm} \times
F(Q^2,P^2)\,
\mathcal{A}(Q)^{\mu \nu \rho \sigma} q_\nu \, \left( g_{\rho\lambda} - \frac{P_\rho P_\lambda}{P^2} \right)
\epsilon_{\sigma\alpha \beta \tau} Q^\alpha q^\beta,
\label{effvertex:D*Dgamma}
\end{eqnarray}
where $\mu$, $\lambda$, and $\tau$ are the Lorentz indices of the virtual photon, $D^{*0}$, and real photon, respectively.



 \end{document}